\documentclass[citeautoscript,floatfix,aps,prx,twocolumn,
      superscriptaddress]{revtex4-2}

\usepackage{graphicx}
\usepackage{dcolumn}
\usepackage{bm}
\usepackage{appendix}
\usepackage{amsmath}
\usepackage{braket}
\usepackage[breaklinks=true,colorlinks,citecolor=blue,linkcolor=blue,urlcolor=blue]{hyperref}
\usepackage{subfigure}
\usepackage{float}

\usepackage{changes}

\newcommand{\pf}{\mathrm{pf}}

\begin{document}


\title{Neural Network-Augmented Pfaffian Wave-functions for \\
Scalable Simulations of Interacting Fermions}

\author{Ao Chen}
\affiliation{Center for Computational Quantum Physics, Flatiron Institute, New York 10010, USA}
\affiliation{Division of Chemistry and Chemical Engineering, California Institute of Technology, Pasadena, California 91125, USA}
\affiliation{Center for Electronic Correlations and Magnetism, University of Augsburg, 86135 Augsburg, Germany}

\author{Zhou-Quan Wan}
\affiliation{Center for Computational Quantum Physics, Flatiron Institute, New York 10010, USA}

\author{Anirvan Sengupta}
\affiliation{Center for Computational Quantum Physics, Flatiron Institute, New York 10010, USA}
\affiliation{Center for Computational Mathematics, Flatiron Institute, 162 5th Avenue, New York, New York 10010, USA}
\affiliation{Department of Physics and Astronomy, Rutgers University, Piscataway, New Jersey 08854, USA}

\author{Antoine Georges}
\affiliation{Coll{\`e}ge de France, 11 place Marcelin Berthelot, 75005 Paris, France}
\affiliation{Center for Computational Quantum Physics, Flatiron Institute, New York 10010, USA}
\affiliation{CPHT, CNRS, {\'E}cole Polytechnique, IP Paris, F-91128 Palaiseau, France}
\affiliation{DQMP, Universit{\'e} de Gen{\`e}ve, 24 quai Ernest Ansermet, CH-1211 Gen{\`e}ve, Suisse}

\author{Christopher Roth}
\affiliation{Center for Computational Quantum Physics, Flatiron Institute, New York 10010, USA}

\date{\today}

\begin{abstract}
Developing accurate numerical methods for strongly interacting fermions is crucial for improving our understanding of various quantum many-body phenomena, especially unconventional superconductivity. Recently, neural quantum states have emerged as a promising approach for studying correlated fermions, highlighted by the hidden fermion and backflow methods, which use neural networks to model corrections to fermionic quasiparticle orbitals.
In this work, we expand these ideas to the space of Pfaffians, a wave-function that naturally expresses superconducting pairings, and propose the hidden fermion Pfaffian state (HFPS), which flexibly represents both unpaired and superconducting phases and scales to large systems with favorable asymptotic complexity. 
In our numerical experiments, HFPS provides state-of-the-art variational accuracy in different regimes of both the attractive and repulsive Hubbard models. We show that the HFPS is able to capture both s-wave and d-wave pairing, and therefore may be a useful tool for modeling phases with unconventional superconductivity.

\end{abstract}
\maketitle

\section{Introduction} \label{sec: introduction}

Understanding the behavior of strongly interacting fermions is a prominent challenge in condensed matter physics.
A remarkable feature of strongly correlated fermionic systems is that they host phases with distinct physical properties 
but very small differences in total energy. 
This is notoriously the case for quantum materials with strong electronic correlations, such as copper-oxide 
(cuprate), high critical temperature superconductors~\cite{bednorz1986possible, Wu_PRL87_cuprate}. 
But, remarkably, it is also a feature of even the most idealized lattice model of interacting fermions,
the Hubbard model. 
It is indeed becoming increasingly apparent that the phase diagram of the Hubbard model hosts a rich diversity of possible 
phases, the actual ground state depending sensitively on details 
of the Hamiltonian parameters ~\cite{Qin_ARCMP22_Hubbard,arovas_ARCMP_2022}.   

This is both an experimental opportunity, allowing for the switching between phases with different properties by slightly altering the environment, 
and a major numerical challenge. 
Computational methods must indeed reach exquisite accuracy in order to reliably predict phase diagrams,  
often on the scale of only a few thousandths of the bare electronic energy scales defining the system. 

Variational wave-function methods are a tool of choice to address this problem at zero temperature \cite{Becca_17_VMCtext,Wu_Science24_Vscore}. 
Because the dimension of the Hilbert space depends exponentially on the size of the system, the full many-body wave-function is an intractable object for moderately sized systems. As a result, the wavefunction must be represented in a compressed form, using a parametrized variational {\it ans\"{a}tze}. 

One of the most efficient compression methods is the fermionic matrix product state representation (fMPS) which, combined with the 
density matrix renormalization group algorithm (DMRG) has transformed our ability to compute the properties 
of (quasi-) one-dimensional systems~\cite{White_PRL92_DMRG, White_PRB93_DMRG, Schollwock_AP11_MPS, White_PRL98_DMRG-tJ, Jiang_Science19_DMRG-Hubbard, jiang2022stripe}. 
Generalization of fMPS to higher dimensions such as fermionic projected entangled-pair states (fPEPS), or generally fermionic tensor networks, provides asymptotically exact solutions for fermionic systems with arbitrary dimensions, but is bottlenecked by huge optimization complexity~\cite{Corboz_PRB09_fTN, Corboz_PRB10_fPEPS, Kraus_PRA10_fPEPS, Mortier_SP25_fTN, Liu_PRL25_Hubbard-fPEPS}.

Quantum Monte Carlo (QMC) methods, including determinant QMC (DQMC)~\cite{Blankenbecler_PRD81_DQMC, Hirsch_PRB85_DQMC-Hubbard, White_PRB89_DQMC-Hubbard, Scalapino_PRB93_DQMC, Assaad_08_DQMC} and auxiliary-field QMC (AFQMC)~\cite{Sugiyama_AP86_AFQMC, Zhang_PRB97_AFQMC, Shi_PRB13_AFQMC}, provide accurate solutions for systems without sign problems but are limited by either exponential costs or systematic bias when the sign problem exists~\cite{Troyer_PRL05_SignProblem}. 
Variational Monte Carlo (VMC) optimization of many-body fermionic wave-functions~\cite{Ceperley_PRB77_fVMC, Yokoyama_JPSJ87_VMC-Hubbard1, Yokoyama_JPSJ87_VMC-Hubbard2, Yokoyama_JPSJ87_VMC-Hubbard3, Gros_PRB88_VMC-tJ, Gros_AP89_fVMC, Paramekanti_PRL01_VMC-Hubbard, Sorella_PRL02_VMC-tJ, Arun_PRB04_VMC-cuprates, Tocchio_PRB08_VMC-Hubbard, Tahara_JPSJ08_mVMC, Misawa_CPC19_mVMC} 
is a powerful method to overcome the limitations of the aforementioned methods, especially within the strongly correlated regime, but its accuracy depends strongly on the specific parametrization of the wave-function and the possible physical biases in 
choosing this parametrization.

Recently, a more `agnostic' parametrization of many-body variational wave-functions has been introduced, 
based on artificial neural networks (ANN)~\cite{Carleo_Science17_NQS}.
Neural quantum states (NQS) are a much more expressive class of variational wave-functions which encode quantum correlations in the `synapses' of an ANN. 
This approach has become a mainstream method for modeling spin systems and has provided great insights into quantum spin liquids~\cite{Nomura_PRX21_PPRBMJ1J2, Roth_PRB23_GCNN, Chen_NP24_MinSR, Viteritti_PRB25_TransformerQSL}. On the other hand, the development of NQS for fermionic systems has progressed somewhat more slowly. 
Although great progress has been made in continuous space for quantum chemistry problems~\cite{Choo_NC20_QuChem, Pfau_PRR20_FermiNet, Hermann_NC20_PauliNet, vonGlehn_arxiv23_Psiformer, Pfau_Science24_ChemExcited} and the electron gas~\cite{Pescia_PRB24_MessagePassing, Smith_PRL24_NQS-ElectronGas}, fermionic lattice problems are still challenging due to strong interactions and correlations in lattice models. Several NQS architectures have been proposed for fermions, including a combination of a Pfaffian and a restricted Boltzmann machine (PP+RBM)~\cite{Nomura_PRB17_PPRBM}, neural network backflow (NNBF)~\cite{Luo_PRL19_Backflow}, and hidden fermion determinant states (HFDS)~\cite{Moreno_PNAS22_HFDS}. These architectures, nevertheless, either use an ANN as a generalized Jastrow factor, which only provides a weak correction to the mean-field wave-function, or have unfavorable scaling complexity. 

Since superconductivity is a focal point of research on strongly correlated systems, it is natural to consider 
variational wave-functions which explicitly embody the possibility for fermions to form pairs, a classic 
example being the mean-field BCS {\it ansatz}~\cite{Bardeen_PR57_BCS}. 
More generally, a mean-field wavefunction with pairing, called a Thouless state \cite{thouless1961vibrational}, forms a class of wave-functions with the structure of a Pfaffian~\cite{coleman1965structure, Bouchaud_JPF88_pfaffian} when it is projected onto a fixed number of particles. In quantum chemistry calculations, this is often called the antisymmetric geminal power (AGP)~\cite{surjan1999introduction}. 
In addition to superconductors, Pfaffians have been shown to describe states with non-abelian anyons~\cite{moore1991nonabelions, storni2010fractional}, and to provide a good representation of the $5/2$ fractional quantum Hall state. Most generally, the Pfaffian parametrizes a broad class of anti-symmetric objects which encompasses Slater determinant states~\cite{Tahara_JPSJ08_mVMC} and paired BCS wave-functions. 
While Pfaffians are in general expensive objects, taking $\mathcal{O}(n^3)$ time to compute, Pfaffian-based VMC can be accelerated by imposing sublattice structure~\cite{Misawa_CPC19_mVMC} and applying low-rank update (LRU) techniques~\cite{Xu_CPC22_LRU}, to efficiently push simulations to large scales. Some early works have considered the use of Pfaffians in the framework of neural quantum states~\cite{Nomura_PRB17_PPRBM, Fore_PRR23_NQS-NeutronStar, Kim_CP24_NQSpfaffian, Gao_arxiv24_NeuralPfaffian, Fore_CP25_NQS-NeutronStar}, but this line of research remains rather uncharted, especially in combination with the most expressive ANN architectures and efficient optimization methods.

In this manuscript, we propose an ANN-augmented Pfaffian variational wave-function, 
which we call hidden fermion Pfaffian states (HFPS), for the simulation of interacting fermions. 
This class of wave-functions, which generalizes  the hidden fermion formalism~\cite{Moreno_PNAS22_HFDS}, provides an 
architecture which is compatible with deep ANNs and large-scale VMC simulations.
We show that the HFPS can successfully approximate ground states of the Hubbard model in various phases, including s-wave superconductors, Mott insulators, and the `stripe' spin and charge-density wave phase. 
In small systems where exact diagonalization (ED) is feasible, or for systems where QMC is sign-free, 
HFPS provide variational energies close to those benchmark methods. 
In all of the studied cases, HFPS gives state-of-the-art variational results, often surpassing other methods by more than an order of magnitude in variational accuracy. This gives promise that HFPS can resolve the small energy differences between competing phases and map out precise phase diagrams of strongly interacting models. Furthermore, HFPS scales rather well with system size and offers the promise of accessing large enough systems to understand fermionic phases with long-range correlations, including uncoventional superconductors.  

\section{Methods}

\begin{figure*}[t]
    \centering
    \includegraphics[width=0.8\linewidth]{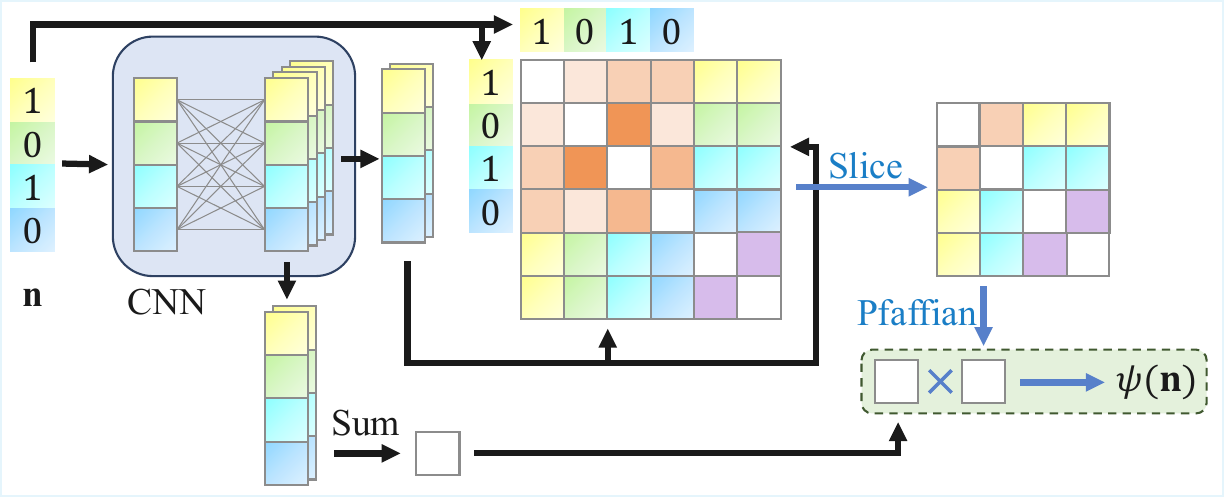}
    \caption{Illustration of HFPS architecture in Eq.\,\eqref{eqn: hfps}. Here we show a one-dimensional (1D) spinless fermion system for simplicity. In this example, there are four orbitals, $M=4$, two visible fermions, $N=2$, and two hidden fermions, $\tilde N = 2$. The CNN outputs are separated into two parts along the channel dimension, one for $\tilde{\mathbf{F}}^{vh}(\mathbf{n})$ and the other for $J(\mathbf{n})$. The $4\times4$ Pfaffian matrix contains four $2\times2$ parts, $\mathbf{n} \star \mathbf{F}^{vv} \star \mathbf{n}$, $\mathbf{n} \star \tilde{\mathbf{F}}^{vh}(\mathbf{n})$, $-\tilde{\mathbf{F}}^{vh}(\mathbf{n})^T \star \mathbf{n}$, and $\tilde{\mathbf{F}}^{hh}$. The orange blocks represent the matrix elements of $\mathbf{F}^{vv}$, and the purple ones represent the matrix elements of $\tilde{\mathbf{F}}^{hh}$. The remaining colored blocks represent $\tilde{\mathbf{F}}^{vh}(\mathbf{n})$ generated by a CNN which are equivariant to a translation of the input Fock state $\mathbf{n}$.}
    \label{fig: HFPS}
\end{figure*}

\subsection{Pfaffian wave-function}

The Pfaffian wave-function is a projected solution of the general bilinear Hamiltonian
\begin{equation} \label{eqn: mf_hamiltonian}
    \hat{\mathcal{H}}_0 = \sum_{p,q}^{M} \left( t_{pq} \hat c_p^\dagger \hat c_q + \frac{1}{2}\Delta_{pq} \hat c_p^\dagger \hat c_q^\dagger + \frac{1}{2} \Delta_{qp}^* \hat c_p \hat c_q \right),
\end{equation}
where the indices $p$ and $q$ run over all orbitals of the system including spatial and spinful degrees of freedom, and $M$ is the total number of orbitals. In Appendix \ref{sec: diagonalization}, we discuss how to obtain the ground state of $\hat{\mathcal{H}}_0$ using the Bogoliubov transformation. 
A common form of the ground state, often known as the Thouless state, can be expressed as
\begin{equation} \label{eqn: pf_exp}
    \ket{\psi_0} = \exp \left(\frac{1}{2} \sum_{p,q}^M F_{pq} \hat c_p^\dagger \hat c_q^\dagger \right) \ket{0},
\end{equation}
where $\mathbf{F}$ is an $M \times M$ anti-symmetric matrix depending on $t_{pq}$ and $\Delta_{pq}$, and $\ket{0}$ denotes the vacuum state. Since the BCS Hamiltonian is a special case of $\hat{\mathcal{H}}_0$, $\ket{\psi_0}$ can express the BCS state by
\begin{equation} \label{eqn: BCS}
\begin{split}
    \ket{\psi_\mathrm{BCS}} &= \prod_\mathbf{k} (u_\mathbf{k} + v_\mathbf{k} \hat c_{\mathbf{k},\uparrow}^\dagger \hat c_{-\mathbf{k},\downarrow}^\dagger) \ket{0} \\
    &= \left(\prod_\mathbf{k} u_\mathbf{k}\right) \times 
    \exp \left(\sum_{i,j} F_{ij} \hat c_{i,\uparrow}^\dagger \hat c_{j,\downarrow}^\dagger \right) \ket{0},
\end{split}
\end{equation}
where $\hat c_{\mathbf{k},\sigma}^\dagger = \sum_j \hat c_{j,\sigma}^\dagger e^{i \mathbf{k \cdot r}_j} / \sqrt{N_\mathrm{site}}$ is the quasi-particle in the momentum space of a lattice with $N_\mathrm{site}$ unit cells, and $F_{ij} = \sum_\mathbf{k} e^{i \mathbf{k}\cdot(\mathbf{r}_i - \mathbf{r}_j)} v_\mathbf{k} / u_\mathbf{k} N_\mathrm{site}$.

For closed systems where the number of particles is conserved, it is convenient to project $\ket{\psi_0}$ onto a sector of fixed particle number
\begin{equation} \label{eqn: pfaffian state}
    \ket{\psi_\pf} = \hat{\mathcal{P}}_{N} \ket{\psi_0}
    = \frac{1}{(N/2)!} \left( \sum_{p<q} F_{pq} \hat c_p^\dagger \hat c_q^\dagger \right)^{N/2} \ket{0},
\end{equation}
where $\hat{\mathcal{P}}_{N}$ is the projector onto the sector with $N$ fermions.
Given a Fock state
\begin{equation}
    \ket{\mathbf{n}} = \prod_p^M (\hat c_p^\dagger)^{n_p} \ket{0},
\end{equation}
where $n_p$ is $0/1$ for an empty/occupied orbital and $\sum_p n_p = N$, the wave-function of $\ket{\psi_\mathrm{pf}}$ is
\begin{equation} \label{eqn: pf_wf}
    \psi_\pf(\mathbf{n}) = \braket{\mathbf{n}|\psi_\pf} =  \textrm{pf} ({\bf n} \star \mathbf{F} \star {\bf n} ),
\end{equation}
where $\mathrm{pf}(\mathbf{A})$ represents the Pfaffian of a matrix $\mathbf{A}$. 
A brief introduction to Pfaffians and to some of their mathematical properties is provided in Appendix \ref{sec: pfaffian}.
The $\star$ symbol represents a slicing operation selecting the rows of $\mathbf{F}$ by $\mathbf{n} \star \mathbf{F}$ and columns by $\mathbf{F} \star \mathbf{n}$ according to the occupied orbitals of ${\bf n}$. The sliced matrix size becomes $N \times N$ if the Fock state $\ket{\mathbf{n}}$ has $N$ particles. Eq.\,\eqref{eqn: pf_wf} is the most common form of the Pfaffian wave-function in VMC, where independent elements of $\mathbf{F}$ are treated as trainable variational parameters.

The Pfaffian wave-function can also express a Slater determinant state.  Utilizing Eq.\,\eqref{eq:pf->det_pf}, one can rewrite a Slater determinant wave-function in Eq.\,\eqref{eqn: slater unprojected} into
\begin{equation} \label{eqn: det_state}
    \psi_\mathrm{det}(\mathbf{n}) = \det (\mathbf{n} \star \mathbf{u})
    = \pf (\mathbf{n} \star \mathbf{uJu}^T \star \mathbf{n}),
\end{equation}
where $\mathbf{u}$ is an $M \times N$ matrix denoting single-particle orbitals in a Slater determinant state, and $\mathbf{J}$ is any anti-symmetric matrix satisfying $\pf \mathbf{J} = 1$. Therefore, a Pfaffian wave-function with $\mathbf{F} = \mathbf{uJu}^T$ expresses a Slater determinant state.

In summary, the Pfaffian wave-function in Eq.\,\eqref{eqn: pf_wf} is a general projected solution of the bilinear Hamiltonian, where the Slater determinant state in Eq.\,\eqref{eqn: det_state} with non-interacting quasi-particles, and the projected BCS state in Eq.\,\eqref{eqn: BCS} with Cooper pairs are special cases. It can also express superconducting states with p-wave and d-wave pairings given suitable $\hat{\mathcal{H}}_0$. Furthermore, the structure of the Pfaffian wave-function allows for fast low-rank updates~\cite{Xu_CPC22_LRU} and sublattice translational symmetry~\cite{Misawa_CPC19_mVMC}. These techniques greatly accelerate simulations in big systems and make large-scale VMC optimization possible.

\subsection{Hidden fermion Pfaffian state}

\begin{figure*}[t]
    \centering
    \includegraphics[width=0.9\linewidth]{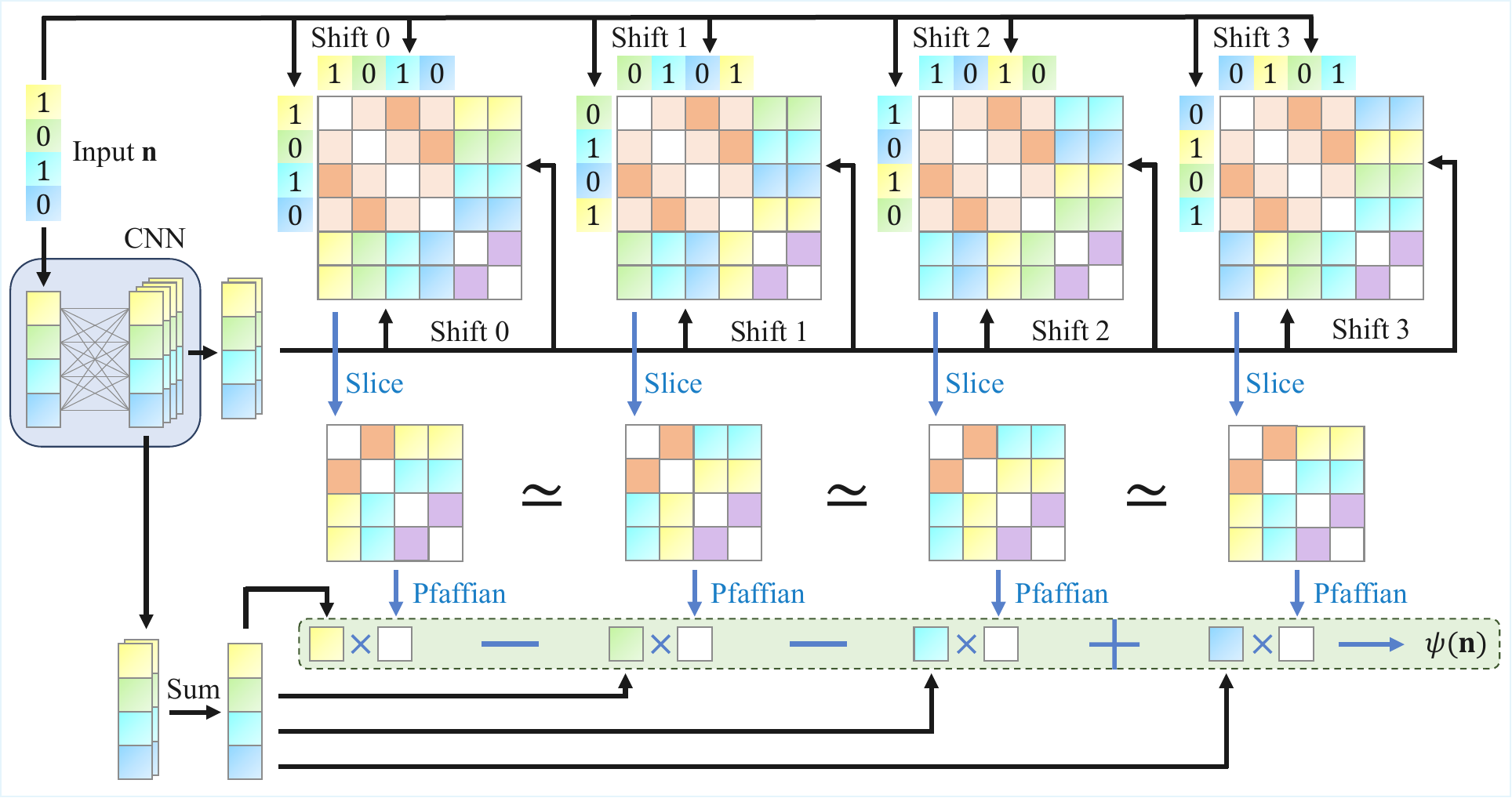}
    \caption{Illustration of HFPS architecture with sublattice and translation symmetry as an extension of the architecture in Fig.\,\ref{fig: HFPS}. The $\mathbf{F}^{vv}$ matrix elements have a banded structure due to the sublattice symmetry discussed in Appendix \ref{sec: symmetry}. The sliced matrices therefore only differ by a permutation applied to both the rows and columns, and the Pfaffian values are equivalent up to a permutation sign, as indicated by the ``$\simeq$'' symbol in the plot. In practice, the Pfaffian is only computed once to reduce the computational cost. Finally, the translation symmetrization in Eq.\,\eqref{eqn: trans_symm} is performed to obtain the HFPS with translation symmetry.}
    \label{fig: HFPS_translation}
\end{figure*}
    
In Ref.\cite{Moreno_PNAS22_HFDS}, it was shown how correlations between particles can be taken into account by 
enlarging the physical Hilbert space with `hidden' degrees of freedom, while keeping a Slater determinant form of the wave-function. Higher-order correlations between electrons emerge from a projection onto the physical Hilbert space. 

Here, we generalize this concept to Pfaffian wave-functions.
We expand the Hilbert space with $M$ orbitals and $N$ visible fermions to include additional ${\tilde M}$ hidden orbitals and ${\tilde N}$ hidden fermions. The Pfaffian state in this expanded Hilbert space, which we name the hidden fermion Pfaffian state (HFPS), becomes
\begin{equation}
\begin{split}
    \ket{\psi} = \frac{1}{(N_\mathrm{tot} / 2)!}
    \Big( &\sum_{p<q} F^{vv}_{p q} \hat c^\dagger_p \hat c^\dagger_q \\
    + &\sum_{\tilde p < \tilde q} F^{hh}_{\tilde p \tilde q} \hat d_{\tilde p}^\dagger \hat d_{\tilde q}^\dagger 
    + \sum_{p,\tilde p} F^{vh}_{p \tilde p} \hat c_p^\dagger \hat d_{\tilde p}^\dagger \Big)^{N_\mathrm{tot} / 2} \ket{0},
\end{split}
\end{equation}
where ${\bf F}^{vv}$, ${\bf F}^{vh}$, and ${\bf F}^{hh}$ are matrices representing the various couplings between visible and hidden fermions, $\hat c_p^\dagger$ and $\hat d_{\tilde p}^\dagger$ denote respectively visible and hidden fermion operators, and $N_\mathrm{tot} = N + {\tilde N}$ is the total number of fermions.
The wave-function component, similar to Eq.\,\eqref{eqn: pf_wf}, is given by
\begin{equation}
\begin{split}
    \psi(\mathbf{n}, \tilde{\mathbf{n}}) &= \braket{\mathbf{n}, \tilde{\mathbf{n}}|\psi} \\
    &= \pf \left[ (\mathbf{n}, \tilde{\mathbf{n}}) \star \begin{pmatrix}
        \mathbf{F}^{vv} & \mathbf{F}^{vh} \\ -(\mathbf{F}^{vh})^T & \mathbf{F}^{hh}
    \end{pmatrix} \star (\mathbf{n}, \tilde{\mathbf{n}}) \right]\\
    &= \pf \begin{pmatrix}
        {\bf n} \star \mathbf{F}^{vv} \star {\bf n} & {\bf n} \star \mathbf{F}^{vh} \star \tilde{\mathbf{n}}\\
        - \tilde{\mathbf{n}} \star (\mathbf{F}^{vh})^T \star {\bf  n} & \tilde{\mathbf{n}} \star \mathbf{F}^{hh} \star \tilde{\mathbf{n}}
    \end{pmatrix}.
\end{split}
\end{equation}
where $\mathbf{n}$ and $\tilde{\mathbf{n}}$ denote the occupation number of visible and hidden fermions, respectively.

In order to describe a physical wave-function, we need to project $\ket{\psi}$ back onto the Hilbert space of the real fermions. A natural choice is to make the hidden fermion occupations $\tilde{\mathbf{n}}$ depend on visible fermions $\mathbf{n}$, i.e. ${\bf \tilde n} \rightarrow {\bf \tilde n}({\bf n})$. Then 
\begin{equation}
    \psi(\mathbf{n}) = \mathrm{pf} \begin{pmatrix}
        {\bf n} \star \mathbf{F}^{vv} \star {\bf n} & {\bf n} \star \mathbf{F}^{vh} \star \tilde{\mathbf{n}} ({\bf n})\\
        - \tilde{\mathbf{n}} ({\bf n}) \star (\mathbf{F}^{vh})^T \star {\bf  n} & \tilde{\mathbf{n}}  ({\bf n}) \star \mathbf{F}^{hh} \star \tilde{\mathbf{n}} ({\bf n})
    \end{pmatrix},
\end{equation}
In practice, one cannot easily parametrize the discrete distribution $\tilde{\mathbf{n}}(\mathbf{n})$ by ANNs. Alternatively, we define an $M \times \tilde{N}$ matrix $\mathbf{F}^{vh}(\mathbf{n}) = \mathbf{F}^{vh} \star \tilde{\mathbf{n}} ({\bf n})$ and an $\tilde{N} \times \tilde{N}$ matrix $\mathbf{F}^{hh}(\mathbf{n}) = \tilde{\mathbf{n}}  ({\bf n}) \star \mathbf{F}^{hh} \star \tilde{\mathbf{n}}  ({\bf n})$, such that
\begin{equation} \label{eqn: raw_hfps}
    \psi(\mathbf{n}) = \mathrm{pf} \begin{pmatrix}
        {\bf n} \star \mathbf{F}^{vv} \star {\bf n} & {\bf n} \star \mathbf{F}^{vh}(\mathbf{n}) \\
        - \mathbf{F}^{vh}(\mathbf{n})^T \star {\bf  n} & \mathbf{F}^{hh}(\mathbf{n})
    \end{pmatrix}.
\end{equation}
Then one can use ANNs to parameterize $\mathbf{F}^{vh}(\mathbf{n})$ and $\mathbf{F}^{hh}(\mathbf{n})$.

In Appendix \ref{sec: HFDS}, we review the HFDS~\cite{Moreno_PNAS22_HFDS}, an earlier work combining the idea of hidden fermions with Slater determinants. The HFDS wave-function, as shown in Eq.\,\eqref{eqn: hfds}, can be written as
\begin{equation} \label{eqn: HFDS}
    \psi_\mathrm{HFDS}(\mathbf{n}) = \det \begin{pmatrix}
        \mathbf{n} \star \mathbf{u}^v \\
        \mathbf{u}^h(\mathbf{n})
    \end{pmatrix},
\end{equation}
where $\mathbf{u}^v$ is the visible orbitals, and $\mathbf{u}^h(\mathbf{n})$ is the hidden orbitals parametrized by ANNs. Similar to Eq.\,\eqref{eqn: det_state}, one can utilize Eq.\,\eqref{eq:pf->det_pf} to rewrite $\psi_\mathrm{HFDS}$ into
\begin{equation} \label{eqn: hfds->hfps}
\begin{split}
    \psi_\mathrm{HFDS}(\mathbf{n}) &= \pf \left[ 
        \begin{pmatrix}
            \mathbf{n} \star \mathbf{u}^{v} \\ 
            \mathbf{u}^{h}(\mathbf{n}) \\
        \end{pmatrix}
        \mathbf{J}
        \begin{pmatrix}
            \mathbf{n} \star \mathbf{u}^{v} \\ 
            \mathbf{u}^{h}(\mathbf{n}) \\
        \end{pmatrix}^T
    \right], \\
\end{split}
\end{equation}
where $\mathbf{J}$ is an arbitrary $(N + \tilde{N}) \times (N + \tilde{N})$ anti-symmetric matrix with $\pf \mathbf{J} = 1$. The HFPS wave-function in Eq.\,\eqref{eqn: raw_hfps} can express any HFDS wave-function in Eq.\,\eqref{eqn: hfds->hfps} by letting $\mathbf{F}^{vv} = \mathbf{u}^v \mathbf{J} (\mathbf{u}^v)^T$, $\mathbf{F}^{vh}(\mathbf{n}) = \mathbf{u}^v \mathbf{J} \mathbf{u}^h(\mathbf{n})^T$, and $\mathbf{F}^{hh}(\mathbf{n}) = \mathbf{u}^h(\mathbf{n}) \mathbf{J} \mathbf{u}^h(\mathbf{n})^T$.
Conversely, not all HFPS can be expressed by HFDS. For instance, a full-rank $\mathbf{F}^{vv}$ cannot be decomposed into $\mathbf{u}^v \mathbf{J} (\mathbf{u}^v)^T$. Therefore, the HFPS can be viewed as a generalization of HFDS.

In most modern ANN architectures, especially those utilized in NQS including the convolutional neural network (CNN)~\cite{Choo_PRB19_J1J2CNN, Liang_MLST23_CNNJ1J2, Chen_NP24_MinSR}, group CNN (GCNN)~\cite{Roth_PRB23_GCNN}, and transformers~\cite{Viteritti_PRL23_Transformer, Chen_arxiv25_CTWF}, the shape of output matrices is usually $N_\mathrm{site} \times C$, where $N_\mathrm{site}$ is the number of sites in the system and $C$ is the number of output channels in the network. As shown by Fig.\,\ref{fig: HFPS}, these outputs can be directly reshaped into the required shape $M \times \tilde{N}$ for the matrix $\mathbf{F}^{vh}(\mathbf{n})$ while keeping translation equivariance. Nevertheless, the network output shape is not compatible with the $\tilde{N} \times \tilde{N}$ shape of $\mathbf{F}^{hh}(\mathbf{n})$. 
To solve this issue, we utilize the spectral theory of anti-symmetric matrices to decompose $\mathbf{F}^{hh}(\mathbf{n})$ into $\mathbf{F}^{hh}(\mathbf{n}) = \mathbf{V}^T(\mathbf{n}) \tilde{\mathbf{F}}^{hh} \mathbf{V}(\mathbf{n})$, where $\tilde{\mathbf{F}}^{hh}$ is an $\tilde{N} \times \tilde{N}$ anti-symmmetric matrix, $\mathbf{V}(\mathbf{n})$ is an $\tilde{N} \times \tilde{N}$ unitary matrix, and we attribute all $\mathbf{n}$ dependence to $\mathbf{V}(\mathbf{n})$ without loss of generality. By defining $\tilde{\mathbf{F}}^{vh}(\mathbf{n}) = \mathbf{F}^{vh}(\mathbf{n}) \mathbf{V}^\dagger(\mathbf{n})$ and $\mathbf{U}(\mathbf{n}) = \mathrm{diag}(\mathbf{1}, \mathbf{V}(\mathbf{n}))$, we rewrite Eq.\,\eqref{eqn: raw_hfps} into
\begin{equation} \label{eqn: hfps}
\begin{split}
    \psi(\mathbf{n}) &= \pf \left[ 
        \mathbf{U}^T(\mathbf{n})
        \begin{pmatrix}
            {\bf n} \star \mathbf{F}^{vv} \star {\bf n} & {\bf n} \star \tilde{\mathbf{F}}^{vh}(\mathbf{n}) \\
            - \tilde{\mathbf{F}}^{vh}(\mathbf{n})^T \star {\bf  n} & \tilde{\mathbf{F}}^{hh}
        \end{pmatrix}
        \mathbf{U}(\mathbf{n})
    \right] \\
    &= J(\mathbf{n}) \times \pf 
        \begin{pmatrix}
            {\bf n} \star \mathbf{F}^{vv} \star {\bf n} & {\bf n} \star \tilde{\mathbf{F}}^{vh}(\mathbf{n}) \\
            - \tilde{\mathbf{F}}^{vh}(\mathbf{n})^T \star {\bf  n} & \tilde{\mathbf{F}}^{hh}
        \end{pmatrix},
\end{split}
\end{equation}
where we have utilized Eq.\,\eqref{eq:pf->det_pf} in the last step and defined $J(\mathbf{n}) = \det(\mathbf{U}(\mathbf{n}))$. Therefore, we drop the dependence of hidden-hidden pairing on $\mathbf{n}$ by introducing an additional Jastrow factor $J(\mathbf{n})$. In practice, $J({\bf n})$ and $\tilde{\mathbf{F}}^{vh} ({\bf n})$ are computed by the ANN, while $\mathbf{F}^{vv}$ and $\tilde{\mathbf{F}}^{hh}$ are directly treated as variational parameters. In Fig.\,\ref{fig: HFPS}, we show how to obtain $J({\bf n})$ and $\tilde{\mathbf{F}}^{vh} ({\bf n})$ from a CNN and combine them with the Pfaffian. In Appendix \ref{sec: HFPS->NNBF}, we further show that the HFPS can be viewed as a Pfaffian NNBF with controlled rank, which allows us to perform low-rank updates (LRU) in Appendix \ref{sec: LRU} to greatly accelerate the computation of Pfaffian matrices. In Fig.\,\ref{fig: HFPS_translation}, we also show an illustration of HFPS with sublattice structure, which reduces the number of pfaffian computations required to build a translationally symmetric wave-function.

\subsection{Optimization} \label{subsec: optimization}

Typically, NQS are trained using stochastic reconfiguration (SR), which implements imaginary time evolution. Negating errors that arise from sampling bias and model expressivity, the variational state $\ket{\psi_\tau}$ at training iteration $\tau$ can be expressed in terms of the initial state $\ket{\psi_0}$ as $\ket{\psi_\tau} = \sum_i e^{-\tau E_i} \langle i | \psi_0 \rangle \ket{i}$,
where $\ket{i}$'s are the eigenstates of the Hamiltonian with energies $E_i$. This implies that convergence time strongly depends on the overlap between the initial state and the ground state. As an additional consequence, a variational state with a pathological initialization, such that the overlap is very large with a different eigenstate than the ground state, is very likely to get stuck in a local minimum. Therefore, we improve training efficiency by starting the mean-field part $\mathbf{F}^{vv}$ of HFPS from a mean-field state that we expect to have sizeable overlap with the ground state. We either optimize a Slater determinant state in Eq.\,\eqref{eqn: det_state}, a BCS state in Eq.\,\eqref{eqn: BCS}, or a full Thouless state in Eq.\,\eqref{eqn: pf_exp}, to obtain an optimal mean-field state $\ket{\psi_0}$ and a suitable initial $\mathbf{F}^{vv}$. We train these models using exact gradient descent of variational energy with automatic differentiation, taking advantage of the fact that observables of mean-field states can be computed in $\mathcal{O}(M^2)$ time without Monte-Carlo sampling.
A well-known problem with mean-field methods is that they tend to overestimate long-range order. On the square lattice Hubbard, it has been shown that a mean-field Slater determinant trained with a small interaction, $U \approx 2.7$ for $1/8$ doping, has the largest overlap with the ground state at $U=8$~\cite{Zheng_Science17_Hubbard}. Therefore, we initialize our time evolution with a mean-field state trained at $|U|=3$. In some cases, we also find it is best to add a d-wave pairing during mean-field optimization, as detailed in \ref{subsec: stripe}.

After properly initializing $\mathbf{F}^{vv}$, we train all variational parameters in Eq.\,\eqref{eqn: hfps}, including $\mathbf{F}^{vv}$, $\tilde{\mathbf{F}}^{hh}$, and ANN weights $\boldsymbol{\theta}$, to search for the ground state of interacting fermion systems by employing SR~\cite{Sorella_PRL98_SR, Sorella_JCP07_SR, Mazzola_JCP12_SR} to perform imaginary-time evolution. To solve the SR equation, we utilize minimum-norm SR (MinSR)~\cite{Chen_NP24_MinSR} and subsampled projected-increment natural gradient descent (SPRING)~\cite{Goldshlager_JCP24_SPRING} with momentum $\mu=0.5$. 

In most simulations, we utilize a CNN similar to the ResNet in Ref.\,\cite{Chen_NP24_MinSR, Chen_arxiv25_CTWF} with 16 layers (8 residual blocks), kernel size $3\times3$, channel number $C=32$, and in total around 150K parameters. Among 32 channels, 28 of them are transformed into $\tilde{\mathbf{F}}^{vh}$ with $\tilde N = 14$ hidden fermions in the studied spinful system, and the remaining channels are used for the generalized Jastrow factor $J(\mathbf{n})$. The symmetry projection explained in Appendix \ref{sec: symmetry} is also implemented to improve the accuracy. For simulations in \ref{subsec: stripe}, we utilize a GCNN~\cite{Roth_PRB23_GCNN} as also described in Appendix \ref{sec: symmetry}. Unless otherwise specified, the total number of samples in each VMC iteration is fixed at $N_s = 16384$, and the training takes around $10^4$ VMC iterations. 

\subsection{Reducing the Asymptotic Complexity}

\begin{figure}[t]
    \centering
    \includegraphics[height=7.5cm]{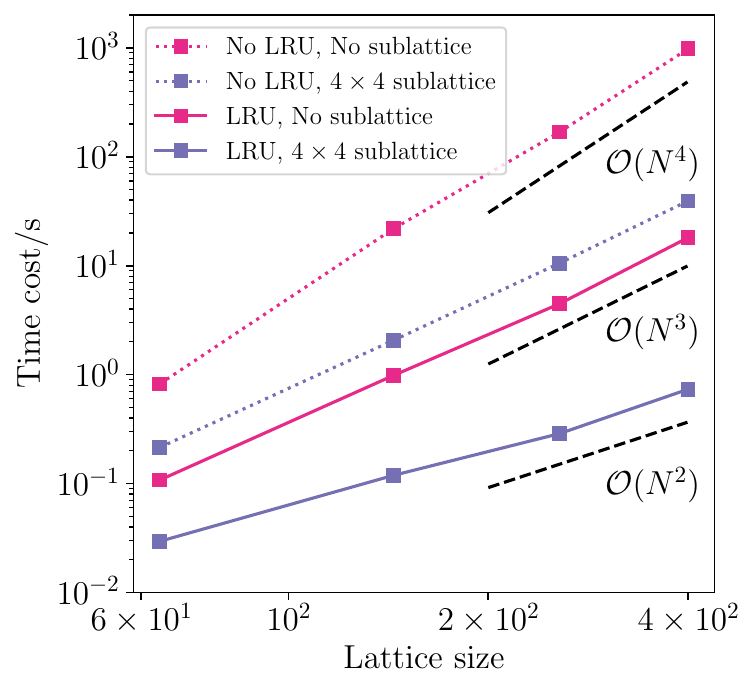}
    \caption{Comparison of the time cost of a HFPS forward pass with and without the acceleration given by low-rank updates (LRU) and sublattice symmetry.}
    \label{fig: time_cost}
\end{figure} 

For large systems, the time complexity of every VMC iteration is dominated by the following parts.
\begin{itemize}
    \item Monte Carlo sampling or local energy computation in VMC: $\mathcal{O}(N)$;
    \item Computing Pfaffian of matrices: $\mathcal{O}((N+\tilde{N})^3)$;
    \item Imposing translation symmetry on Pfaffian: $\mathcal{O}(M)$.
\end{itemize}

{
\renewcommand{\arraystretch}{1.5}
\begin{table}[b]
    \centering
    \begin{tabular}{c|c|c||c|c}
        NQS & LRU & Sublattice & Forward pass & Full VMC \\
        \hline\hline
        PP+RBM~\cite{Nomura_PRB17_PPRBM} & Yes & Yes & $\mathcal{O}(N^2)$ & $\mathcal{O}(N^3)$ \\
        \hline
        NNBF~\cite{Luo_PRL19_Backflow} & No & No & $\mathcal{O}(N^4)$ & $\mathcal{O}(N^5)$ \\
        \hline
        HFDS~\cite{Moreno_PNAS22_HFDS} & Yes & No & $\mathcal{O}(N^3)$ & $\mathcal{O}(N^4)$ \\
        \hline
        HFPS & Yes & Yes & $\mathcal{O}(N^2)$ & $\mathcal{O}(N^3)$ \\
    \end{tabular}
    \caption{Complexity of different types of fermionic NQSs with translation symmetry}
    \label{tab: NQS_complexity}
\end{table}
}

Then the total complexity is $\mathcal{O}(N M (N + \tilde{N})^3)$ for each VMC step. This complexity exceeds that of a CNN or GCNN, which costs $\mathcal{O}(M)$ per forward pass and $\mathcal{O}(N M)$ in each VMC step. The particle density $n$ and hidden fermion number $\tilde N$ are usually kept unchanged when one increases the system size, so the complexity can be simplified as $\mathcal{O}(N^5)$. This complexity becomes the main bottleneck in the NQS simulation of large fermion systems, limiting most of the previous fermionic NQS attempts to small systems or low accuracy~\cite{Luo_PRL19_Backflow, Moreno_PNAS22_HFDS, Romero_CP25_NNBF}.

\begin{figure}[t]
    \centering
    \includegraphics[height=7.5cm]{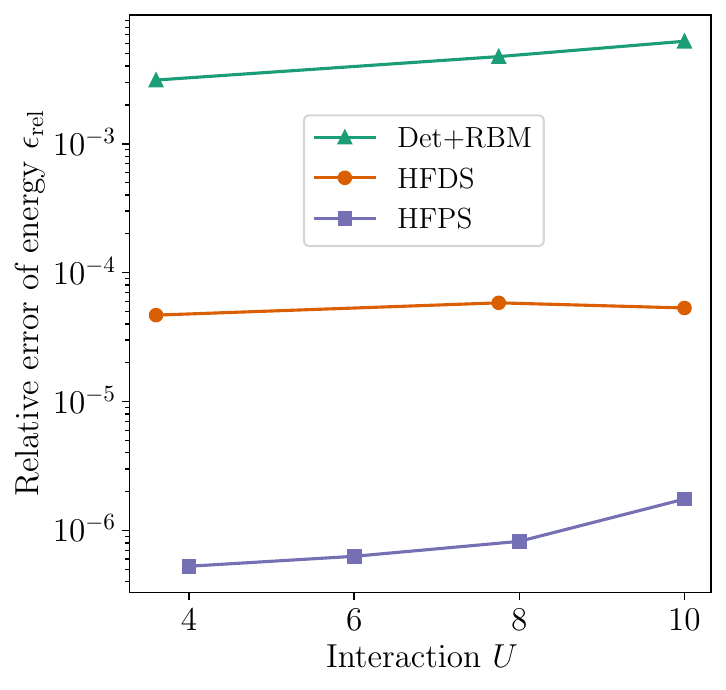}
    \caption{Relative error of energy $\epsilon_\mathrm{rel}$ in the $4 \times 4$ Hubbard model with particle density $n=5/8$. The definition of $\epsilon_\mathrm{rel}$ is given in Eq.\,\eqref{eqn: rel_error}, and the error is computed with respect to the ED energy. The results given by the determinant with RBM (Det+RBM) and hidden fermion determinant state (HFDS)~\cite{Moreno_PNAS22_HFDS} are provided to compare with our HFPS energy.}
    \label{fig: 4x4_FL}
\end{figure}

Fortunately, there are several techniques one can employ in HFPS to greatly reduce the complexity. In Monte Carlo sampling or local energy computation, the sliced Pfaffian orbitals are often generated by replacing a few rows and columns from previous Pfaffian orbitals if the Hamiltonian only contains local interactions. In this case, one does not have to recompute the full Pfaffian, as the low-rank update (LRU) in Appendix \ref{sec: LRU} can be utilized to reduce the complexity from $\mathcal{O}((N+\tilde{N})^3)$ to $\mathcal{O}(N^2 \tilde{N})$. Furthermore, one can also employ sublattice structure in Appendix \ref{sec: symmetry} so that the computational cost of translation symmetry projection does not scale up with $M$. As illustrated in Fig.\,\ref{fig: HFPS_translation}, the computation of the translation-symmetrized wave-function involves $M$ terms, but they are equivalent up to a fermion permutation sign due to sublattice structure, so one only needs to compute $S$ terms, where $S$ is the size of the sublattice unit cell independent of the full system size. With these techniques, the time complexity of the HFPS forward pass is reduced to $\mathcal{O}(N^2)$ and the complexity of each VMC step is reduced to $\mathcal{O}(N^3)$.

In Fig.\,\ref{fig: time_cost}, we show the time cost of $10^3$ parallel HFPS forward pass on an A6000 GPU. As illustrated in the figure, the complexity is greatly reduced from $\mathcal{O}(N^4)$ to $\mathcal{O}(N^2)$ by utilizing LRU and sublattice. For the largest lattice with 400 sites, the forward pass is accelerated by a factor of $10^3$, making it possible to simulate large fermionic systems.

\begin{figure*}[t]
    \centering
    \subfigure{
        \includegraphics[height=6cm]{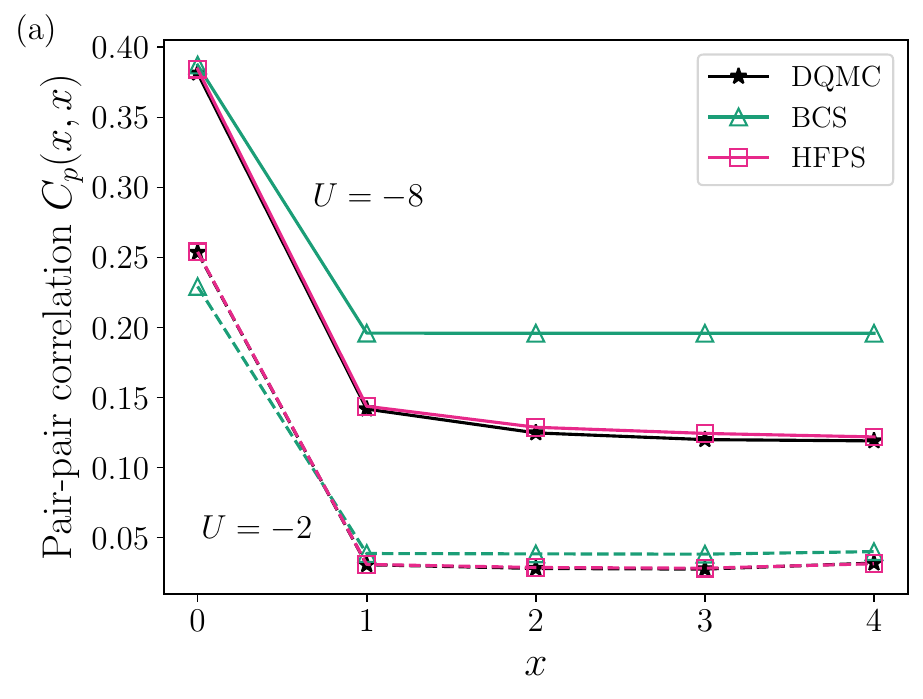}
    }
    \subfigure{
        \includegraphics[height=6cm]{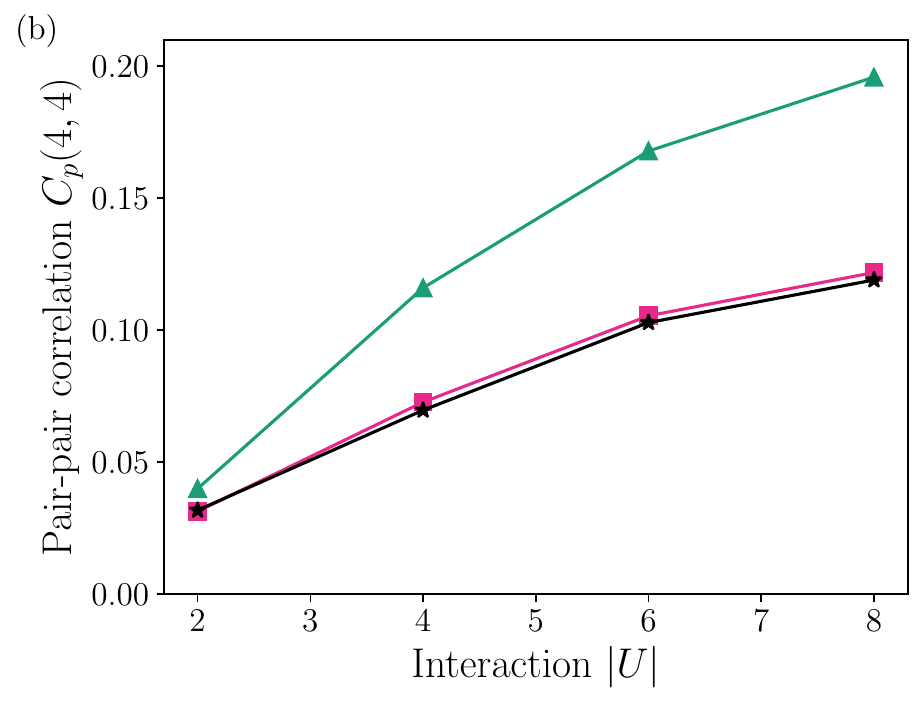}
    }
    \caption{The pair-pair correlation $C_p(\mathbf{r})$ in the $8 \times 8$ Hubbard model with attractive interaction $U<0$ and particle density $n=7/8$. The definition of $C_p(\mathbf{r})$ is given in Eq.\,\eqref{eqn: pair-pair}. The methods shown here include the BCS wave-function in Eq.\,\eqref{eqn: BCS} with minimized variational energy, our HFPS, and the determinant quantum Monte Carlo (DQMC) for an exact reference in this sign-free system. (a) $C_p(\mathbf{r})$ is measured at $\mathbf{r} = (x, x)$. The dashed and solid lines represent $U=-2$ and $U=-8$, respectively. (b) $C_p(\mathbf{r})$ is measured at $\mathbf{r} = (4, 4)$, the furthest distance in the $8\times8$ system with PBC. The uncertainty is smaller than the marker size in both panels.}
    \label{fig: BCS-BEC}
\end{figure*}

Although the LRU and sublattice techniques can be utilized in HFPS to provide great acceleration, they have not been utilizied in previous fermionic NQS work. As shown in Tab.\,\ref{tab: NQS_complexity}, only PP+RBM~\cite{Nomura_PRB17_PPRBM}, which is a relatively weak correction on the mean-field Pfaffian state, has utilized these techniques, while NNBF~\cite{Luo_PRL19_Backflow} and HFDS~\cite{Moreno_PNAS22_HFDS} have not. Therefore, we expect HFPS to be an ideal NQS architecture for large fermionic systems with strongly correlated electrons.

\section{Results}

To illustrate the performance of HFPS, we employ it to study the two-dimensional Hubbard model on the square lattice
\begin{equation} \label{eqn: Hubbard}
    \hat{\mathcal{H}} = -t \sum_{\braket{ij},\sigma} \hat c_{i,\sigma}^\dagger \hat c_{j,\sigma} + U \sum_i \hat n_{i,\uparrow} \hat n_{i,\downarrow}.
\end{equation}
We will show the performance of HFPS in different phases of the Hubbard model. To quantify the accuracy of variational states, we measure the variational energy $E$ in VMC and compute the relative error of the variational energy defined as
\begin{equation} \label{eqn: rel_error}
    \epsilon_\mathrm{rel} = \frac{E - E_0}{E_\infty - E_0},
\end{equation}
where $E_0$ is the exact ground-state energy, usually provided by ED in small systems or QMC in sign-free cases, and $E_\infty$ is the energy zero point given by the energy at infinite temperature. The reason of introducing $E_\infty$ is to remove the dependency of $\epsilon_\mathrm{rel}$ on a constant shift of Hamiltonian~\cite{Wu_Science24_Vscore}.

The raw data of our numerical experiments is included in Appendix \ref{sec: raw data}.



\subsection{Small-scale benchmark}

We first consider a case in which the density is low (close to quarter filling).  
Electrons are then relatively free to move around in the Hubbard model, 
and the system is in a rather weakly correlated regime even if $U$ is large.
In the thermodynamic limit, the system is expected to be in a Fermi liquid state with well-defined quasiparticle excitations.
As a starting point, we study the performance of HFPS in this simple case to demonstrate its ability to encode correlations. 

In Fig.\,\ref{fig: 4x4_FL}, we show the results of various fermionic NQSs in the $4 \times 4$ Hubbard model with periodic boundary conditions (PBC), visible particle number $N = 10$, and particle density $n=5/8$, within the Fermi liquid regime. The HFPS provides a great improvement in variational energy compared with the neural Jastrow method and HFDS~\cite{Moreno_PNAS22_HFDS}. Compared with the best previous NQS result given by HFDS, the HFPS error is roughly reduced by a factor of 100. We attribute this improvement to the utilization of a modern deep network architecture in HFPS as compared to the shallow fully-connected network in HFDS. We note that it is awkward to implement an HFDS with expressive architectures, including CNNs and transformers, as the required shape $\tilde N \times (M + \tilde N)$ of matrix $\mathbf{u}^h(\mathbf{n})$ in Eq.\,\eqref{eqn: HFDS} is incompatible with the usual network output shape.

\subsection{Superconductivity with attractive interactions} \label{sec: attractive sc}

When the on-site interaction of the Hubbard model is attractive with $U<0$, the electrons form Cooper pairs and exhibit s-wave superconductivity. 
This superconducting state evolves from a BCS regime~\cite{Bardeen_PR57_BCS} at smaller $|U|$ to a strong coupling regime at larger $|U|$ where local pairs form with a large binding energy but become phase coherent at a smaller energy scale, forming a Bose-Einstein condensate (BEC)~\cite{Chen_PhysRep05_BCS-BEC}.
Although the nature of the off-diagonal long-range order (ODLRO) in the ground-state is qualitatively captured by the BCS wave-function in Eq.\,\eqref{eqn: BCS}, the BCS state becomes quantitatively inaccurate at moderate to large $|U|$.
To quantify the superconductivity in a system with a conserved particle number, the usual order parameter $\braket{\hat \Delta(\mathbf{r})} = \braket{\hat c_{\mathbf{r}, \uparrow} \, \hat c_{\mathbf{r}, \downarrow}}$ in the BCS wave-function is not directly accessible. Instead, we define the pair-pair correlation
\begin{equation} \label{eqn: pair-pair}
    C_p(\mathbf{r}) = \braket{\hat \Delta^\dagger(\mathbf{r}) \hat \Delta(\mathbf{0})}.
\end{equation}
Although $C_p(\mathbf{r})$ remains finite even without Cooper pairs when $|\mathbf{r}|$ is small, it can serve as an order parameter of BCS superconductivity when $|\mathbf{r}| \to \infty$, corresponding to ODLRO. 
Therefore, we can use the long-range pair correlations as a proxy for the pairing amplitude in the thermodynamic limit 
\begin{equation} \label{eqn: ODLRO}
|\langle {\hat \Delta} \rangle| \sim \left|\lim_{|{\bf r}| \rightarrow \infty} C_p(\mathbf{r}) \right|^{1/2}.
\end{equation}

In Fig.\,\ref{fig: BCS-BEC}, we show $C_p(\mathbf{r})$ given by several methods in the $8 \times 8$ attractive Hubbard model with particle density $n = 7/8$ PBC. The BCS state is defined in Eq.\,\eqref{eqn: BCS} and optimized by exact gradient descent of energy. It is a qualitatively correct description, as we expect for attractive electrons, and provides good accuracy under weak interaction. However, the BCS wave-function tends to overestimate the long-range correlation when the interaction is strong because of the underestimation of quantum fluctuations. Consequently, it is not a quantitatively accurate description for electrons with strong attractions. In HFPS, although the mean-field part $\mathbf{F}^{vv}$ is initialized by the BCS wave-function, it is trained to encode correlations through the ANN and achieves great accuracy in predicting pair-pair correlations, within the errorbar of the exact result given by DQMC. The relative error of energy $\epsilon_\mathrm{rel}$ given by HFPS also reaches the level of $10^{-4}$, greatly outperforming the $10^{-2}$ level provided by the BCS wave-function. Therefore, HFPS provides a reliable approach in studying the behavior of systems with BCS superconductivity and strong interactions, for instance, the crossover to BEC.

\subsection{Mott insulator}

At half-filling (one particle per site on average) and for strong repulsive interactions $U>0$, the kinetic energy of electrons is suppressed, and the ground-state of the Hubbard model on the square lattice is a Mott insulator with 
antiferromagnetic long-range order.
On a bipartite lattice, the half-filling case is special since it does not display a sign problem in Monte Carlo methods and 
projective auxiliary field quantum Monte Carlo (AFQMC) becomes exact. 
Hence, AFQMC provides a benchmark to which variational wave-functions can be compared.
Furthermore, the spin-balanced Hubbard model with $U$ and $-U$ can be considered equivalent at half filling. 
This can be shown from a partial particle-hole transformation on spin-down operators $\hat c_{i, \downarrow} \to \mathrm{sign}(i) \,\hat c_{i,\downarrow}^\dagger$, where $\mathrm{sign}(i) = +1$ for one bipartite lattice and $-1$ for the other. Then the original Hubbard Hamiltonian in Eq.\,\eqref{eqn: Hubbard} is transformed into
\begin{equation} \label{eqn: particle_hole_transform}
    \hat{\mathcal{H}} = -t \sum_{\braket{ij},\sigma} \hat c_{i,\sigma}^\dagger \hat c_{j,\sigma} - U \sum_i \hat n_{i,\uparrow} \hat n_{i,\downarrow} + U \sum_i \hat n_{i,\uparrow}.
\end{equation}
As the last term is a constant for a system with conserved spin-up and spin-down particles, this transformation shows that the attractive and repulsive Hubbard models are equivalent at half filling. In this work, we utilize this freedom of basis choice to simulate an attractive Hubbard model for the Mott insulator phase in HFPS, which allows us to use the same technique as the simulations in \ref{sec: attractive sc} and initialize HFPS using a BCS state in Eq.\,\eqref{eqn: BCS} trained with exact gradient.

\begin{figure}[t]
    \centering
    \includegraphics[height=7.5cm]{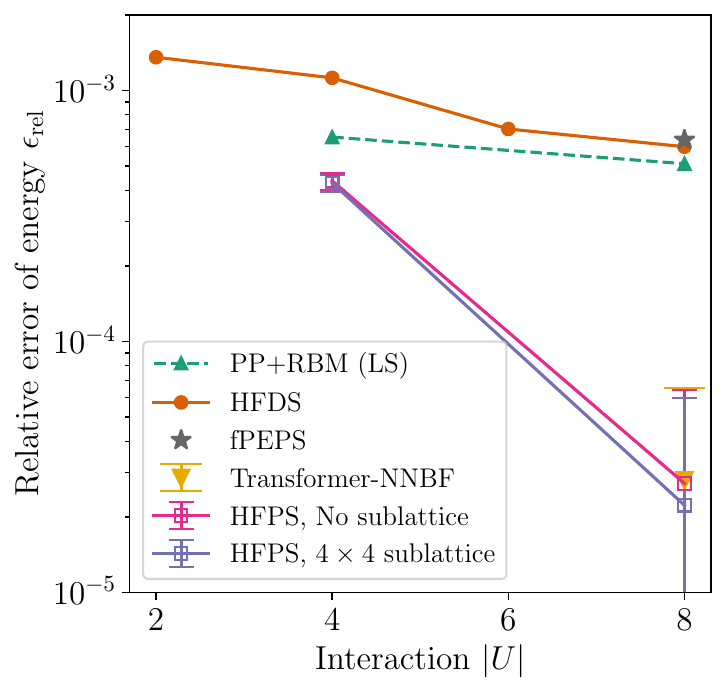}
    \caption{Relative error of energy $\epsilon_\mathrm{rel}$ in the $8 \times 8$ Hubbard model at half-filling (particle density $n=1$). Here we include the results provided by PP+RBM with a Lanczos step (LS)~\cite{Nomura_PRB17_PPRBM}, HFDS~\cite{Moreno_PNAS22_HFDS}, Transformer-NNBF~\cite{Gu_arxiv25_NQS-Hubbard}, and fPEPS with bond dimension $D=16$ (provided by J. Gray, private communication) for comparison. The ground-state energy is estimated by AFQMC at $U=-4$ and $-8$~\cite{Shi_PRA15_HubbardAFQMC, Wu_Science24_Vscore} for the computation of $\epsilon_\mathrm{rel}$. The uncertainty of the relative error of HFPS and Transformer-NNBF mainly comes from the uncertainty of AFQMC, and in other methods the uncertainty is smaller than the marker size.
    }
    \label{fig: half-filling}
\end{figure}

In Fig.\,\ref{fig: half-filling}, we show the energy error of different NQSs in the $8 \times 8$ lattice at half-filling. The HFDS result is provided in the system with PBC along one side and anti-periodic boundary condition (APBC) along the other, while other results are provided with PBC along both sides. The AFQMC reference energy is selected with respective boundary conditions for each method to compute the relative error $\epsilon_\mathrm{rel}$. The existing numerical methods, including PP+RBM with an exact Lanczos step (LS), HFDS~\cite{Moreno_PNAS22_HFDS}, and fPEPS with bond dimension $D=16$, provide $\epsilon_\mathrm{rel}$ at the level of $10^{-3}$, while HFPS provides better accuracy. The fPEPS result presented here is optimized by the simple update method and not pushed to its limiting expressive power. We expect the gradient-based optimization~\cite{Liu_PRB17_GO-PEPS, Liu_PRB21_GO-PEPS, Liu_PRL25_Hubbard-fPEPS} will reduce the error of fPEPS by a factor of 2 to 10, but HFPS still achieves a similar or better accuracy. Compared to a recent result provided by Transformer-NNBF~\cite{Gu_arxiv25_NQS-Hubbard}, our HFPS has better accuracy. It is worth noting that the uncertainty of the benchmark AFQMC energy is around $10^{-4}$, which leads to a big uncertainty of $\epsilon_\mathrm{rel}$ when the accuracy of HFPS also reaches this level, as shown in Fig.\,\ref{fig: half-filling}.

As also presented in Fig.\,\ref{fig: half-filling}, the HFPS wave-functions with no sublattice and $4 \times 4$ sublattice show similar error because a $4 \times 4$ unit cell is large enough to encode the translational symmetry breaking of the best mean-field state. The sublattice symmetry provides great acceleration in VMC, allowing us to train with more iterations given a similar total time cost. Therefore, the HFPS with sublattice reaches slightly better accuracy compared to the one without sublattice, although the former in principle has lower expressive power.

\begin{figure}[t]
    \centering
    \includegraphics[height=7.5cm]{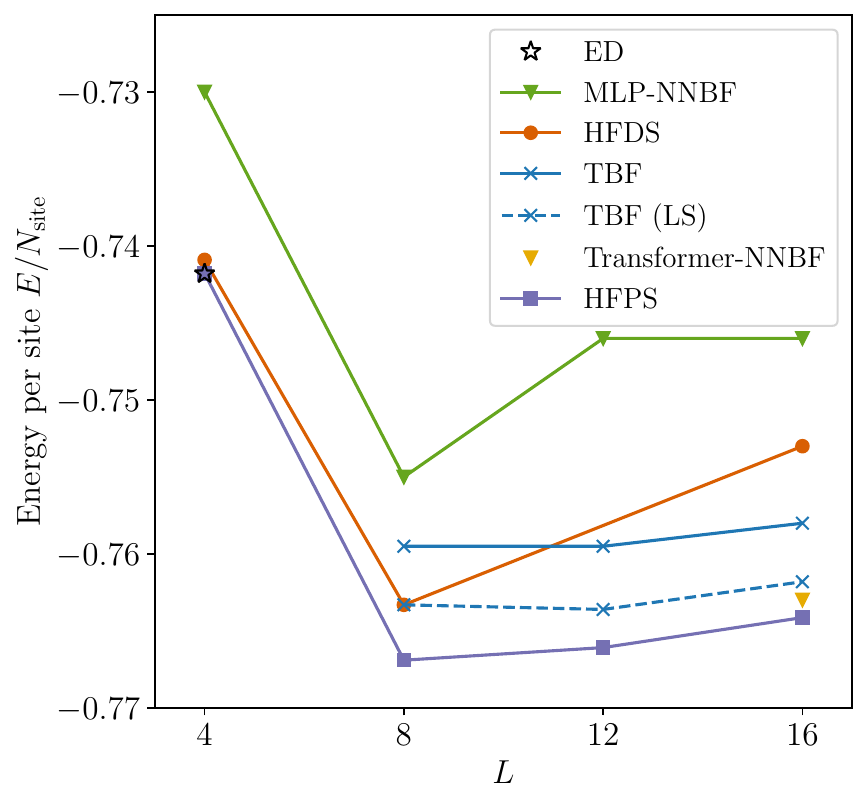}
    
    \caption{Variational energy in the spin- and charge-ordered (stripe) ground state of the $L \times 4$ Hubbard model with repulsive interaction $U=8$ and particle density $n=7/8$. The presented variational results include multi-layer perceptron neural network backflow (MLP-NNBF)~\cite{Luo_PRL19_Backflow}, hidden fermion determinant state (HFDS)~\cite{Moreno_PNAS22_HFDS}, tensor backflow (TBF)~\cite{Zhou_PRB24_TBF}, TBF with a Lanczos step (LS)~\cite{Zhou_PRB24_TBF}, Transformer-NNBF~\cite{Gu_arxiv25_NQS-Hubbard}, and HFPS.}
    \label{fig: CDW}
\end{figure}

\subsection{Stripe phase} \label{subsec: stripe}

\begin{figure*}[t]
    \centering
    \subfigure{
        \includegraphics[width=0.63\textwidth]{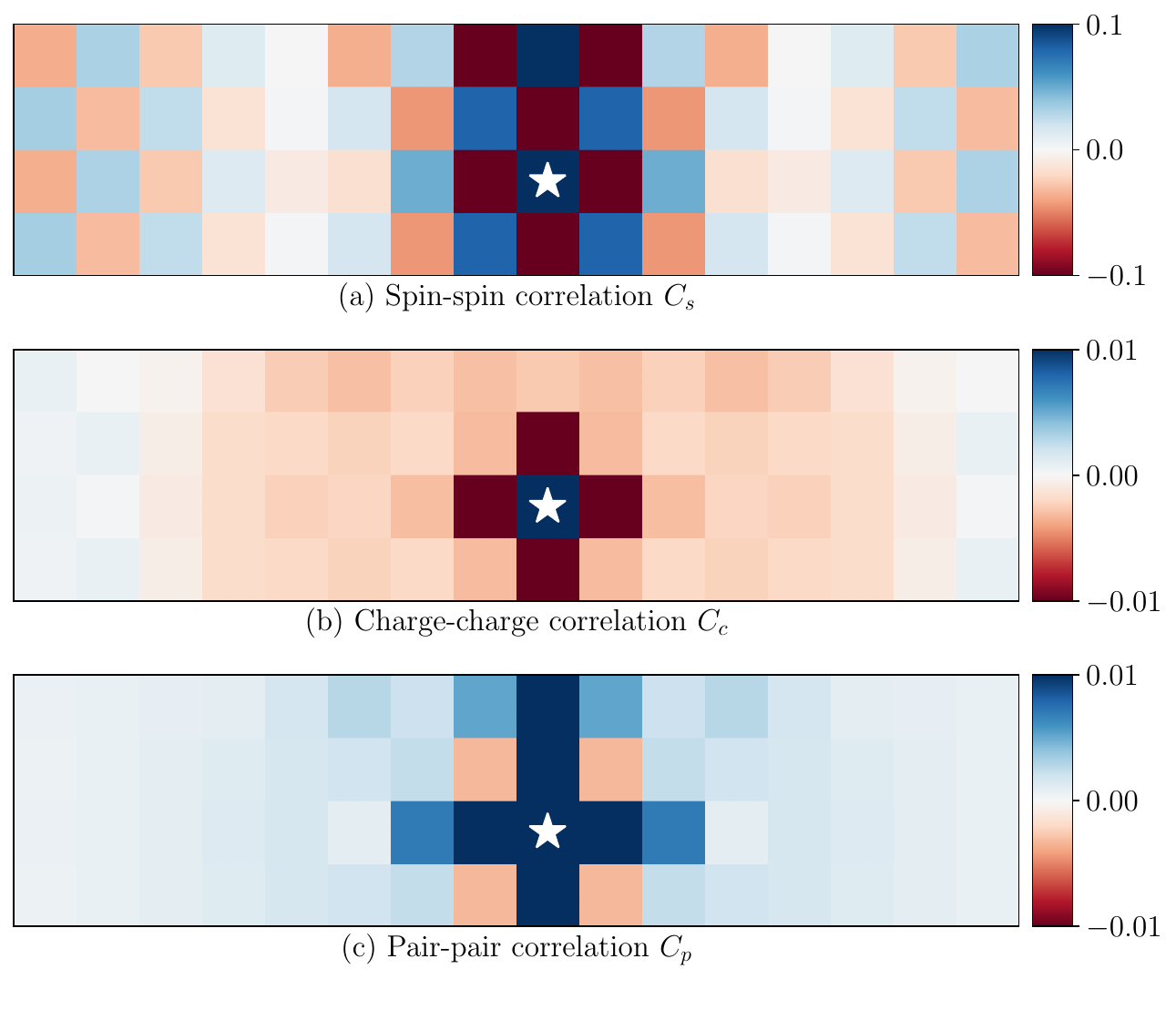}
    }
    \hspace{-0.5cm}
    \subfigure{
        \includegraphics[width=0.35\textwidth]{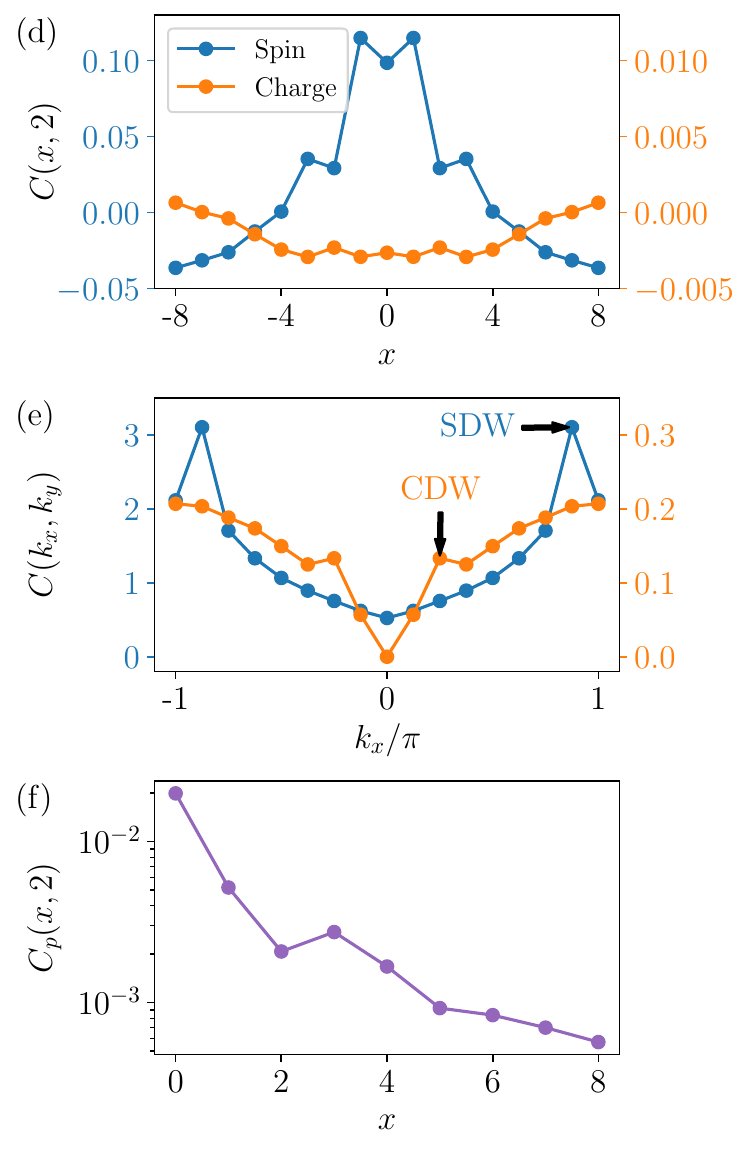}
    }
    \caption{The spin, charge, and pair correlations in the $16 \times 4$ Hubbard model with $1/8$ doping and $U=8$. 
    (a-c) The correlations $C_s$, $C_c$, and $C_p$ defined by Eq.\,\eqref{eqn: spin_correlation}-\eqref{eqn: d-wave pair-correlation} in the full real space. For visibility of weak correlations, their strength is truncated to ceiling values $|C_s| \le 0.1$ and $|C_c|,|C_p| \le 0.01$. The star marks the central site with $\mathbf{r} = (0, 0)$.
    (d) The staggered spin-spin correlation $(-1)^x C_s(\mathbf{r})$ and the charge-charge correlation $C_c(\mathbf{r})$ with $\mathbf{r} = (x, 2)$. The two curves are given different scales to make the oscillations prominent. 
    (e) The spin structure factor $C_s(\mathbf{k})$ with $\mathbf{k} = (k_x, \pi)$ and the charge structure factor $C_c(\mathbf{k})$ with $\mathbf{k} = (k_x, 0)$ in Eq.\,\eqref{eqn: structure_factor}. The arrows indicate the expected structure factor peaks for SDW and CDW. 
    (f) The d-wave pair-pair correlation $C_p(\mathbf{r})$ with $\mathbf{r}=(x, 2)$. The uncertainty is smaller than the marker size in (d-f).
    }
    \label{fig: stripe_correlation}
\end{figure*}

In order to test our method on a computationally challenging task, we study the Hubbard model with small $\delta=1/8$ hole doping (particle density $n=7/8$) and strong repulsive interactions $U=8$. 
In this regime, the Hubbard model displays stripe order, where one-dimensional hole-rich domain walls lie between patches of anti-ferromagnetic order. The sublattice polarization switches across the domain walls so that when electrons hop across, the magnetism is not frustrated. This manifests itself as a density wave pattern of charge and spin ordering, with an associated structure factor peak. At the same time, the doped square lattice Hubbard model has a tendency towards d-wave pairing, and states with off-diagonal long-range order (ODLRO) are energetically competetive~\cite{LeBlanc_PRX15_Hubbard, Zheng_Science17_Hubbard, Qin_PRX20_Hubbard, Xu_Science24_Hubbard}. Determining whether density-wave order, superconducting order, or both, survive to the thermodynamic limit is an ongoing problem for condensed matter physics that requires accurate variational methods. In existing studies, it is widely believed that the ground state of the Hubbard model with $U=8$ and $\delta = 1/8$ is a density wave phase without ODLRO~\cite{Qin_PRX20_Hubbard, wietek2022cooper, Sorella_PRB23_VAFQMC}. While our simulations corroborate these findings, we find that d-wave pairing still plays an essential role in the short-range physics and energetics.

As explained in \ref{subsec: optimization}, suitable initialization with a mean-field state is crucial for avoiding local minima, especially in the stripe regime where there are many competing low-energy states.
For a Hubbard model with $U \ge 0$, the mean-field state with the lowest energy is a Slater determinant state without pairing. For the $4 \times 4$ and $8 \times 4$ lattices, we indeed achieve good accuracy starting from a Slater determinant trained at $U=3$. Nevertheless, we find that on the $16 \times 4$ lattice we are able to improve our final energies by training the mean-field state with an additional d-wave pairing field, i.e.
\begin{equation} \label{eqn: mean field ham}
    \hat{\mathcal{H}}' = \hat{\mathcal{H}} + \Delta \sum_{\braket{ij}} \mathrm{sign}(i,j) (\hat c^\dagger_{i,\uparrow} \hat c^\dagger_{j,\downarrow} - \hat c^\dagger_{i,\downarrow} \hat c^\dagger_{j,\uparrow} + H.C.),
\end{equation}
where $\hat{\mathcal{H}}$ is the original Hubbard Hamiltonian in Eq.\,\eqref{eqn: Hubbard} with a weaker interaction $U=3$, $\Delta$ is the strength of the d-wave field, and $\mathrm{sign}(i,j)$ denotes the sign for d-wave pairing, which is $+1$ for horizontal neighbors and $-1$ for vertical neighbors. The lowest energy state of this Hamiltonian is a general Pfaffian state with both stripe order and d-wave pairing. For the $16 \times 4$ stripe, we train our mean-field state with $\Delta=0.2$. In the following, we utilize a GCNN in our HFPS with roughly 600K parameters.

The variational energies of different methods on $L \times 4$ PBC lattices are presented in Fig.\,\ref{fig: CDW}. In this system, HFPS and the recent Transformer-NNBF~\cite{Gu_arxiv25_NQS-Hubbard} greatly outperform earlier works based on shallow ANNs.
These results showcase the importance of large-scale simulations in fermionic NQS, as also demonstrated in previous simulations on spin systems~\cite{Rende_CP24_SRt, Chen_NP24_MinSR}. Compared to the ground state energy marked by the black star in the small $4\times4$ system, HFPS provides nearly exact energy with $\epsilon_\mathrm{rel} = 1.2\times10^{-5}$, while HFDS has $\epsilon_\mathrm{rel} = 3\times10^{-4}$. In larger systems, we also present tensor backflow (TBF)~\cite{Zhou_PRB24_TBF}, a simple backflow with coefficients directly represented by tensors instead of ANNs, for reference. As shown in the plot, the previous MLP-NNBF~\cite{Luo_PRL19_Backflow} and HFDS~\cite{Moreno_PNAS22_HFDS} provide worse energy than the trivial tensor representation of backflow, indicating that the ANNs in these methods probably do not encode correlations between electrons properly. Our HFPS method, in contrast to these NQSs, outperforms TBF and even the TBF with a Lanczos step (LS), demonstrating that the scalable HFPS architecture captures the essential degrees of freedom in strongly correlated electrons. The energy we obtain in the $16 \times 4$ lattice is $E/N = -0.76413$, outperforming a recent result $E/N = -0.76298$ produced by Transformer-NNBF~\cite{Gu_arxiv25_NQS-Hubbard}. We believe that this energy discrepancy is likely due to the presence of strong short-ranged d-wave pairing in the ground state~\cite{wietek2022cooper}, as we find a fairly similar energy, $E/N = -0.76256$, to Transformer-NNBF when we initialize without a pairing field, which drives us to a meta-stable state without pairing. This is evidence that using Pfaffian-based wave-functions, which are better equipped to model pairing, may be useful even for phases without ODLRO, as long as there is short-range pairing.

Utilizing the accurate numerical solution in the $16 \times 4$ lattice, we compute the correlations in the system to verify the existence of stripe patterns and short-range d-wave pairings, including the spin-spin correlation
\begin{equation} \label{eqn: spin_correlation}
    C_s(\mathbf{r}) = \braket{\hat{\mathbf{S}}_\mathbf{r} \cdot \hat{\mathbf{S}}_\mathbf{0}}
    - \braket{\hat{\mathbf{S}}_\mathbf{r}} \cdot \braket{\hat{\mathbf{S}}_\mathbf{0}},
\end{equation}
the charge-charge correlation
\begin{equation} \label{eqn: charge_correlation}
    C_c(\mathbf{r}) = \braket{\hat{n}_\mathbf{r} \cdot \hat{n}_\mathbf{0}}
    - \braket{\hat{n}_\mathbf{r}} \cdot \braket{\hat{n}_\mathbf{0}},
\end{equation}
and the pair-pair correlation 
\begin{equation} \label{eqn: d-wave pair-correlation}
    C_p(\mathbf{r}) = \braket{\hat \Delta^\dagger(\mathbf{r}) \hat \Delta(\mathbf{0})}
\end{equation}
for d-wave symmetrized pairs
\begin{equation}
    \hat \Delta(\mathbf{r}) = \frac{1}{4} \sum_{\boldsymbol{\delta}} \frac{1}{\sqrt{2}} \mathrm{sign}(\boldsymbol{\delta}) 
    (\hat c_{\mathbf{r},\uparrow} \hat c_{\mathbf{r}+\boldsymbol{\delta}, \downarrow}
    - \hat c_{\mathbf{r},\downarrow} \hat c_{\mathbf{r}+\boldsymbol{\delta}, \uparrow}),
\end{equation}
where $\sum_{\boldsymbol{\delta}}$ iterates over nearest neighbors, and $\mathrm{sign}(\boldsymbol{\delta}) = +1$ for horizontal neighbors and $-1$ for vertical neighbors. The correlations are presented in Fig.\,\ref{fig: stripe_correlation}(a-c).

To show a more clear tendency, in Fig.\,\ref{fig: stripe_correlation}(d) we choose $\mathbf{r} = (x, 2)$ and plot the staggered spin correlation $(-1)^x \, C_s(\mathbf{r})$ and the charge correlation $C_c(\mathbf{r})$ as a function of $x$. The spin correlation shows a strong spin density wave (SDW) with spatial period 16, which matches the $1/8$ doping in the system. The charge correlation indicates a superposition of short-range repulsion and the charge density wave (CDW) with spatial period 8, also consistent with the $1/8$ doping. In Fig.\,\ref{fig: stripe_correlation}(e), we show the spin and charge structure factor
\begin{equation} \label{eqn: structure_factor}
    C_\gamma(\mathbf{k}) = \sum_\mathbf{r} C_\gamma(\mathbf{r}) e^{-i \mathbf{k \cdot r}},
\end{equation}
where $\gamma$ represents either the spin structure $s$ or the charge structure $c$. The momentum cuts are chosen to be $\mathbf{k} = (k_x, \pi)$ for spin structures and $\mathbf{k} = (k_x, 0)$ for charge structures so that they pass through their respective peaks at $\mathbf{k}=(7\pi/8, 0)$ and $\mathbf{k}=(\pi/4, 0)$, reflecting the existence of SDW with period 16 and CDW with period 8.
These results confirm the existence of density wave order in the doped Hubbard model.
In Fig.\,\ref{fig: stripe_correlation}(f), we plot $C_p(\mathbf{r})$ for $\mathbf{r}=(x, 2)$, which shows finite short-range d-wave pairings coexisting with SDW and CDW. Since the pairing decays over long distances, our results at $1/8$-doping do not display long-range superconducting order. However, some mechanisms, e.g., the next-nearest-neighbor hopping $t'$~\cite{Xu_Science24_Hubbard}, might enhance the pairing to turn the system into an unconventional superconductor.

\section{Discussion}

In this work, we introduced an ANN-augmented Pfaffian wave-function, HFPS, 
for performing variational Monte Carlo on interacting fermion problems. 
This construction generalizes the hidden-fermion formalism~\cite{Moreno_PNAS22_HFDS} 
to an anti-symmetrized wave-function of fermionic pairs, the Pfaffian. The motivation for this switch is to accurately model superconductors, as the Pfaffian can represent them at the mean-field level.

We describe how to initialize the HFPS with a good fermionic mean-field, so that the training converges well. Additionally, we utilize various numerical techniques that reduce the overall VMC complexity from $\mathcal{O}(N^5)$ to $\mathcal{O}(N^3)$, hence providing a systematic approach to large-scale simulations of strongly correlated electrons. 

We demonstrate the performance of HFPS in various phases of the Hubbard model, 
including the superconducting phase of the attractive case in both weak coupling (BCS) and strong coupling (BEC),
the antiferromagnetic Mott insulator, and the stripe phase. 
The HFPS is able to achieve significantly higher accuracy than existing variational methods and produces accurate energies and correlation functions in sign-free cases when benchmark comparisons with QMC are possible.

The progress allowed by HFPS pushes fermionic NQS studies from 
proof-of-principle results on small systems to a new stage where it can be used to study challenging problems in a systematic fashion. We believe that this will 
allow us to shed light on fermion problems with strong and long-range correlations in 2D, similar to how NQS has improved our understanding of quantum spin systems.

Finally, we list several possible future directions for HFPS and more 
generally for fermionic NQS.

\begin{itemize}
    \item In Fig.\,\ref{fig: BCS-BEC}, we show that HFPS correctly captures s-wave superconductivity in the attractive Hubbard model and provides systematic improvement on the BCS wave-function. As the Pfaffian wave-function represents a general Gaussian state, it can also capture unconventional pairing (e.g., p-wave and d-wave superconductivity) given a suitable initialization specified by the mean-field Hamiltonian $\hat{\mathcal{H}}_0$ in Eq.\,\eqref{eqn: mf_hamiltonian}. Furthermore, as shown by Fig.\,\ref{fig: CDW} and Fig.\,\ref{fig: stripe_correlation}, the HFPS reaches high accuracy in the stripe phase with strong correlations, which has significant d-wave short-range pairing, and may be closely related to superconducting phases. These results reveal the great potential of HFPS for studying high-temperature superconductivity in the $t$-$t'$ square lattice Hubbard model. While there is strong evidence for d-wave superconductivity, especially with $|t'| > 0$ ~\cite{Qin_PRX20_Hubbard, Sorella_PRB23_VAFQMC, Xu_Science24_Hubbard}, NQS could provide detailed descriptions of these phases, including determining whether there are residual density wave orders, and the spatial extent of the pair binding. 

    \item HFPS could be used to study multi-band/multi-orbital models of 
    strongly correlated electrons, as relevant to the description of many quantum materials such as nickelate superconductors in the two-band case. It could also provide more accurate computational descriptions of cuprates by including the oxygen orbitals. Additionally, HFPS could be used to study Hofstadter-Hubbard models and effective models for moir\'e materials. 

    \item By measuring particle correlations in the many-body NQS, 
    one might be able to construct the effective low-energy Hamiltonian governing the behavior of quasi-particles. This would provide new insights into the mechanism of d-wave pairing in high-temperature superconductors and help us to better understand the rich phases of cuprates and other quantum materials with strong electronic correlations.
    
\end{itemize}

\section*{Acknowledgements}

The ED is performed with \texttt{QuSpin}~\cite{Weinberg_SPP17_QuSpinI, Weinberg_SPP19_QuSpinII}. The NQS simulations are performed by \texttt{Quantax}~\cite{quantax}, in which the scalable ANNs are implemented by \texttt{JAX}~\cite{jax2018github} and \texttt{Equinox}~\cite{Kidger_arxiv21_Equinox}, and the LRU is implemented by \texttt{lrux}~\cite{lrux}. 
A.C. acknowledges Markus Heyl for his support and Johnnie Gray for providing fPEPS data. 
We are grateful to Miguel Morales, Javier Robledo-Moreno, Conor Smith, Shiwei Zhang, and Garnet Chan for useful discussions. The Flatiron Institute is a division of the Simons Foundation.

\appendix

\section{Diagonalization of bilinear Hamiltonian} \label{sec: diagonalization}

The general form of the bilinear Hamiltonian in Eq.\,\eqref{eqn: mf_hamiltonian} can be rewritten as
\begin{equation}
\begin{split}
    \hat{\mathcal{H}}_0 &= \sum_{p,q}^{M} \left( t_{pq} \hat c_p^\dagger \hat c_q + \frac{1}{2}\Delta_{pq} \hat c_p^\dagger \hat c_q^\dagger + \frac{1}{2} \Delta_{qp}^* \hat c_p \hat c_q \right) \\
    &= \frac{1}{2} (\hat{\mathbf{c}}^\dagger, \hat{\mathbf{c}}) \mathbf{H}_0
    \begin{pmatrix}
        \hat{\mathbf{c}} \\ \hat{\mathbf{c}}^\dagger
    \end{pmatrix} + \mathrm{const.},
\end{split}
\end{equation}
where $\hat{\mathbf{c}}^\dagger = (\hat c_1^\dagger, ..., \hat c_M^\dagger)$, and
\begin{equation}
    \mathbf{H}_0 =
    \begin{pmatrix}
        \mathbf{t} & \Delta \\ \Delta^\dagger & -\mathbf{t}^T
    \end{pmatrix}.
\end{equation}
To diagonalize $\hat{\mathcal{H}}_0$, we employ the Bogoliubov transformation
\begin{equation}
    \begin{pmatrix}
        \hat{\boldsymbol{\gamma}} \\ \hat{\boldsymbol{\gamma}}^\dagger
    \end{pmatrix}
    = \mathbf{w}^\dagger
    \begin{pmatrix}
        \hat{\mathbf{c}} \\ \hat{\mathbf{c}}^\dagger
    \end{pmatrix}
    = \begin{pmatrix}
        \mathbf{u}^\dagger & \mathbf{v}^\dagger \\ \mathbf{v}^T & \mathbf{u}^T 
    \end{pmatrix}
    \begin{pmatrix}
        \hat{\mathbf{c}} \\ \hat{\mathbf{c}}^\dagger
    \end{pmatrix},
\end{equation}
where
\begin{equation}
    \mathbf{w} = \begin{pmatrix}
        \mathbf{u} & \mathbf{v}^* \\ \mathbf{v} & \mathbf{u}^*
    \end{pmatrix}
\end{equation}
is the transformation matrix. The fermion anti-commutation relations $\{\hat\gamma_\alpha, \hat\gamma_\beta^\dagger\} = \delta_{\alpha\beta}$ and $\{\hat\gamma_\alpha, \hat\gamma_\beta\} = 0$ imply the constraint $\mathbf{w^\dagger w = \mathbf{1}}$, or equivalently $\mathbf{u}^\dagger \mathbf{u} + \mathbf{v}^\dagger \mathbf{v} = \mathbf{1}$ and $\mathbf{u}^T \mathbf{v} + \mathbf{v}^T\mathbf{u} = \mathbf{0}$. Then $\hat{\mathcal{H}}_0$ becomes
\begin{equation}
    \hat{\mathcal{H}}_0 = \frac{1}{2} 
    (\hat{\boldsymbol{\gamma}}^\dagger, \hat{\boldsymbol{\gamma}}) 
    \mathbf{w}^\dagger \mathbf{H}_0 \mathbf{w}
    \begin{pmatrix}
        \hat{\boldsymbol{\gamma}} \\ \hat{\boldsymbol{\gamma}}^\dagger
    \end{pmatrix}
    + \mathrm{const.}
\end{equation}
By choosing suitable $\mathbf{w}$ that diagonalizes $\mathbf{w}^\dagger \mathbf{H}_0 \mathbf{w}$, we obtain $\hat{\mathcal{H}}_0 = \sum_\alpha (E_\alpha \hat\gamma_\alpha^\dagger \hat\gamma_\alpha + E_\alpha' \hat\gamma_\alpha \hat \gamma_\alpha^\dagger) /2 + \mathrm{const}$.
Due to the particle-hole symmetry in $\hat{\mathcal{H}}_0$, we have $E_\alpha' = -E_\alpha$, and
\begin{equation}
    \hat{\mathcal{H}}_0 = \sum_\alpha^M E_\alpha \hat\gamma_\alpha^\dagger \hat\gamma_\alpha + \mathrm{const.},
\end{equation}
where we assume $E_\alpha \geq 0$, because for any $E_\alpha < 0$ one can perform a particle-hole transformation $\hat \gamma_\alpha^\dagger \to \hat \gamma_\alpha$ to obtain the same form with positive quasi-particle energies.

The ground state $\ket{\psi_0}$ of $\hat{\mathcal{H}}_0$ should satisfy $\hat\gamma_\alpha \ket{\psi_0} = 0$ for all $\alpha$.  A common form is the Thouless state
\begin{equation}
    \ket{\psi_0} = \exp \left(\frac{1}{2} \sum_{p,q}^M F_{pq} \hat c_p^\dagger \hat c_q^\dagger \right) \ket{0},
\end{equation}
where $\mathbf{F} = (\mathbf{vu}^{-1})^*$ is an $M \times M$ anti-symmetric matrix, and $\ket{0}$ denotes the true vacuum state. One can directly verify that
\begin{equation}
    \hat\gamma_\alpha \ket{\psi_0}
    = \sum_p (u_{p\alpha}^* \hat c_p + v_{p\alpha}^* \hat c_p^\dagger) \ket{\psi_0} = 0.
\end{equation}
Therefore, apart from the special case of non-invertible $\mathbf{u}$ which corresponds to the existence of unpaired single-particle orbitals, $\ket{\psi_0}$ indeed represents the ground state solution of $\hat{\mathcal{H}}_0$.

\section{Pfaffian} \label{sec: pfaffian}

The Pfaffian maps a $2n \times 2n$ anti-symmetric matrix $\mathbf{A}$ to a number $\pf \,\mathbf{A}$. The Pfaffian can be formally defined as follows. Partition all numbers $\{1, ..., 2n \}$ into $n$ pairs $\alpha = \{(i_1, j_1), ..., (i_n, j_n)\}$ with $i_k < j_k$ and $i_1<i_2<...<i_n$, in total $(2n-1)!!$ possible partitions. Then the Pfaffian of matrix $\mathbf{A}$ with elements $A_{i,j}$ is given by
\begin{equation} \label{eq:pf_def}
    \pf \mathbf{A} = \sum_{\alpha} \mathrm{sign}(\alpha) \prod_{k=1}^n A_{i_k, j_k},
\end{equation}
where $\mathrm{sign}(\alpha)$ is the parity of the permutation $(i_1, j_1, i_2, j_2,..., i_n, j_n)$. For example,
\begin{equation}
    \pf \begin{pmatrix}
        0 & a \\ -a & 0
    \end{pmatrix} = a,
\end{equation}
\begin{equation}
    \pf \begin{pmatrix}
        0 & a & b & c \\
        -a & 0 & d & e \\
        -b & -d & 0 & f \\
        -c & -e & -f & 0
    \end{pmatrix} = af - be + cd.
\end{equation}
The time complexity of Pfaffian is $\mathcal{O}(n^3)$, the same as determinant. In Appendix \ref{sec: LRU}, we will discuss how to reduce the complexity by employing low-rank updates.

For reference, here we list several important properties of the Pfaffian without proof.
\begin{equation} \label{eq:pf2=det}
        \pf^2(\mathbf{A}) = \det(\mathbf{A}),
    \end{equation}
\begin{equation} \label{eq:pf->det_pf}
    \pf (\mathbf{BAB}^T) = \det(\mathbf{B}) \pf(\mathbf{A}),
\end{equation}
\begin{equation} \label{eq:pf_offdiagonal}
    \pf \begin{pmatrix}
        \mathbf{0} & \mathbf{A} \\
        \mathbf{-A}^T & \mathbf{0}
    \end{pmatrix}
    = (-1)^{n(n-1)/2} \det(\mathbf{A}),
\end{equation}
\begin{equation}  \label{eq:pf_diagonal}
    \pf \begin{pmatrix}
        \mathbf{A} & \mathbf{0} \\ \mathbf{0} & \mathbf{A}'
    \end{pmatrix}
    = \pf(\mathbf{A}) \pf(\mathbf{A}'),
\end{equation}
\begin{equation} \label{eq:pf_low_rank}
    \frac{\pf(\mathbf{A} + \mathbf{BCB}^T)}{\pf(\mathbf{A})}
    = \frac{\pf(\mathbf{C}^{-1} + \mathbf{B}^T \mathbf{A}^{-1} \mathbf{B})}{\pf(\mathbf{C}^{-1})},
\end{equation}
\begin{equation} \label{eq:pf_block}
\begin{split}
    \pf\begin{pmatrix}
        \mathbf{M} & \mathbf{Q} \\
        -\mathbf{Q}^T & \mathbf{N}
    \end{pmatrix} 
    &= \pf(\mathbf{M}) \pf(\mathbf{N} + \mathbf{Q}^T \mathbf{M}^{-1} \mathbf{Q}) \\
    &= \pf(\mathbf{N}) \pf(\mathbf{M} + \mathbf{Q} \mathbf{N}^{-1} \mathbf{Q}^T),
\end{split}
\end{equation}

\begin{equation}
    \frac{\partial \pf \mathbf{A}}{\partial A_{ij}} = \frac{1}{2} \pf(\mathbf{A}) (\mathbf{A}^{-1})_{ji}
\end{equation}

\section{Hidden fermion determinant state} \label{sec: HFDS}

In this section, we review the hidden fermion determinant state (HFDS)~\cite{Moreno_PNAS22_HFDS} using the notations of this paper.

\subsection{Determinant state}

Consider a bilinear Hamiltonian with a conserved number of electrons
\begin{equation}
    \hat{\mathcal{H}}_0 = \sum_{p,q}^M t_{pq} \hat c^\dagger_p \hat c_q,
\end{equation}
where the indices $p$ and $q$ run over all $M$ orbitals of the system, including sites and spins. The single-particle eigenstates of this Hamiltonian define a quasi-particle transformation
\begin{equation} \label{eqn: quasiparticle}
    \hat \gamma^\dagger_{\alpha} = \sum_{p}^M u_{p \alpha} \hat c^\dagger_p.
\end{equation}
The Slater determinant state with $N$ fermions is created by filling the band with $N$ quasiparticles
\begin{equation} \label{eqn: slater unprojected}
    \ket{\psi_\mathrm{det}} = \prod_{\alpha}^{N} \gamma^\dagger_{\alpha} \ket{0},
\end{equation}
whose wave-function component is
\begin{equation}
    \psi_\mathrm{det}(\mathbf{n}) = \braket{\mathbf{n}| \psi_\mathrm{det}} = \det (\mathbf{n} \star \mathbf{u}),
\end{equation}
where $\mathbf{u}$ is an $M \times N$ matrix defined in Eq.\,\eqref{eqn: quasiparticle}. The $\star$ symbol is the same slicing operator used in Eq.\,\eqref{eqn: pf_wf}. As $\ket{\mathbf{n}}$ contains $N$ particles, the shape of the sliced matrix $\mathbf{n} \star \mathbf{u}$ is $N \times N$.

\subsection{Adding hidden fermions}

The HFDS generalizes the Slater determinant to an enlarged Hilbert space with $M + {\tilde M}$ orbitals and $N + {\tilde N}$ fermions. The new quasiparticles are then defined as
\begin{equation}
    \hat \gamma_\alpha^\dagger = \sum_p u^{v}_{p \alpha} \hat c^\dagger_p +   \sum_{\tilde p} u^{h}_{\tilde{p} \alpha} \hat d^\dagger_{\tilde p}
\end{equation}
where $\mathbf{u}^{v}$ and $\mathbf{u}^{h}$ denote the transformation matrices, and $\hat d^\dagger_{\tilde p}$ denote hidden particles. Then the wave-function component of HFDS on a Fock state $\ket{\mathbf{n}, \tilde{\mathbf{n}}}$ is given by
\begin{equation}
    \psi(\mathbf{n}, \tilde{\mathbf{n}}) = \bra{{\bf n }, {\tilde {\bf n}}}  \psi \rangle 
    = \det \begin{pmatrix}
        \mathbf{n} \star \mathbf{u}^{v} \\ 
        \tilde{\mathbf{n}} \star \mathbf{u}^{h} \\
    \end{pmatrix}.
\end{equation}

In order to project $\ket{\psi}$ back onto the physical Hilbert space, we make $\tilde{\mathbf{n}}$ depend on $\mathbf{n}$. Equivalently, one can define configuration-dependent quasiparticle orbitals $\mathbf{u}^{h}(\mathbf{n}) = \tilde{\mathbf{n}}(\mathbf{n}) \star \mathbf{u}^{h}$ such that
\begin{equation} \label{eqn: hfds}
    \psi(\mathbf{n}) 
    = \det \begin{pmatrix}
        \mathbf{n} \star \mathbf{u}^{v} \\ 
        \mathbf{u}^{h}(\mathbf{n}) \\
    \end{pmatrix}.
\end{equation}
In seminal work, the hidden quasiparticle orbitals $\mathbf{u}^h(\mathbf{n})$ are parametrized by ANNs. We show in Eq.\,\eqref{eqn: hfds->hfps} that HFPS is a generalization of HFDS.

\section{HFPS to backflow} \label{sec: HFPS->NNBF}

Before this work, it has been shown that HFDS and NNBF formulations can be converted into each other~\cite{Liu_PRB24_HFDS=NNBF}. Here, we also show that HFPS can be converted into NNBF acting on Pfaffians. A Pfaffian backflow can take several different forms~\cite{Gao_arxiv24_NeuralPfaffian, Kim_CP24_NQSpfaffian}. A possible form is
\begin{equation} \label{eqn: NNBF}
    \psi^{\mathrm{NNBF}}(\mathbf{n}) = \tilde J (\mathbf{n}) \times \pf[ \mathbf{n} \star (\mathbf{F}^0 + \mathbf{F}'(\mathbf{n})) \star \mathbf{n}], 
\end{equation}
where $\tilde J(\mathbf{n})$ is a neural-Jastrow factor, $\mathbf{F}^0$ is the mean-field Pfaffian matrix, and $\mathbf{F}'(\mathbf{n})$ is the backflow generated by ANNs.

On the other hand, by utilizing the block Pfaffian formula in Eq.\,\eqref{eq:pf_block}, one can rewrite the HFPS wave-function in Eq.\,\eqref{eqn: hfps} into
\begin{equation}
\begin{split}
    \psi(\mathbf{n}) &= J(\mathbf{n}) \pf (\tilde{\mathbf{F}}^{hh}) \\
    & \times \pf \left[ \mathbf{n} \star (\mathbf{F}^{vv} + \tilde{\mathbf{F}}^{vh}(\mathbf{n}) (\tilde{\mathbf{F}}^{hh})^{-1} \tilde{\mathbf{F}}^{vh}(\mathbf{n})^T) \star \mathbf{n} \right].
\end{split}
\end{equation}
Compared to the NNBF wave-function $\psi^\mathrm{NNBF}(\mathbf{n})$ in Eq.\,\eqref{eqn: NNBF}, we have $\tilde{J}(\mathbf{n}) = J(\mathbf{n}) \pf (\tilde{\mathbf{F}}^{hh})$ as the neural Jastrow factor, $\mathbf{F}^0 = \mathbf{F}^{vv}$ still given by the non-interacting part, and the backflow
$\mathbf{F}'(\mathbf{n}) = \tilde{\mathbf{F}}^{vh}(\mathbf{n}) (\tilde{\mathbf{F}}^{hh})^{-1} \tilde{\mathbf{F}}^{vh}(\mathbf{n})^T$.

A major difference in the current HFPS formulation is that the backflow matrix $\mathbf{F}'(\mathbf{n})$ now has a controlled rank equal to the number of hidden particles $\tilde{N}$. Therefore, the HFPS provides a physical way to implement low-rank backflow. As explained in Appendix \ref{sec: LRU}, it allows us to perform LRU in VMC to reduce the time cost of computing Pfaffian wave-functions from $\mathcal{O}(N^3)$ to $\mathcal{O}(N^2 \tilde{N})$. By controlling the number of hidden fermions $\tilde{N}$, one can reach a balance between accuracy and efficiency.

\section{Low-rank update} \label{sec: LRU}

In VMC and many other applications, one only alters a few rows and columns in the full Pfaffian matrix each time. Consequently, the new Pfaffian can be computed as a low-rank update (LRU) on the previous Pfaffian to reduce the time complexity from $\mathcal{O}(n^3)$ to $\mathcal{O}(n^2)$. In this section, we will introduce the details of LRU. One can also read Ref.\,\cite{Becca_17_VMCtext, Chen2025} for details.

\subsection{Rank-1 update}
Consider two Fock states $\mathbf{n}$ and $\mathbf{n}_0$ only different by one fermion hopping, and $\mathbf{M} = \mathbf{n} \star \mathbf{F} \star \mathbf{n}$ and $\mathbf{M}_0 = \mathbf{n}_0 \star \mathbf{F} \star \mathbf{n}_0$ only different at the $j$'th row and $j$'th column, then
\begin{equation}
\begin{split}
    \mathbf{M - M}_0 &=
    \begin{pmatrix}
        0 & ... & 0 & -u_1 & 0 & ... & 0 \\ 
        \vdots && \vdots & \vdots & \vdots && \vdots \\ 
        0 & ... & 0 & -u_{j-1} & 0 & ... & 0 \\ 
        u_1 & ... & u_{j-1} & 0 & u_{j+1} & ... & u_{N_e} \\
        0 & ... & 0 & -u_{j+1} & 0 & ... & 0 \\ 
        \vdots && \vdots & \vdots & \vdots && \vdots \\ 
        0 & ... & 0 & -u_{N_e} & 0 & ... & 0 \\
    \end{pmatrix} \\
    &= -\begin{pmatrix}
        u_1 & 0 \\ \vdots & \vdots \\ u_{j-1} & 0 \\ u_j & 1 \\ u_{j+1} & 0 \\ \vdots & \vdots \\ u_{N_e} & 0
    \end{pmatrix}
    \begin{pmatrix}
        0 & 1 \\
        -1 & 0
    \end{pmatrix}
    \begin{pmatrix}
        u_1 & 0 \\ \vdots & \vdots \\ u_{j-1} & 0 \\ u_j & 1 \\ u_{j+1} & 0 \\ \vdots & \vdots \\ u_{N_e} & 0
    \end{pmatrix}^T \\
    &= -\mathbf{v} \mathbf{J}_1 \mathbf{v}^T,
\end{split}
\end{equation}
where $\mathbf{v} = (\mathbf{u}, \mathbf{e})$ is an $N \times 2$ matrix, $\mathbf{u}^T$ is the non-zero row of $\mathbf{M-M}_0$, $\mathbf{e}$ is a one-hot vector with $e_i = \delta_{i,j}$, and
\begin{equation} \label{eq:pfa_identity}
    \mathbf{J}_k = 
    \begin{pmatrix}
        \mathbf{0} & \mathbf{1}_k \\
        -\mathbf{1}_k & \mathbf{0}
    \end{pmatrix}
\end{equation}
is a $2k \times 2k$ anti-symmetric identity matrix with $\mathrm{pf}\mathbf{J} = (-1)^{k(k-1)/2}$ from Eq.\,\eqref{eq:pf_offdiagonal}. By utilizing Eq.\,\eqref{eq:pf_low_rank}, we have
\begin{equation} \label{eq:LRU}
\begin{split}
    \pf \mathbf{M} &= \pf (\mathbf{M}_0 - \mathbf{v} \mathbf{J}_1 \mathbf{v}^T) \\
    &= \frac{\pf \mathbf{M}_0}{\pf (-\mathbf{J}_1^{-1})}
    \pf ( -\mathbf{J}_1^{-1} + 
    \mathbf{v}^T \mathbf{M}^{-1}_0 \mathbf{v}) \\
    &= \pf\mathbf{M}_0 \, \pf \mathbf{R},
\end{split}
\end{equation}
where we have utilized $\mathbf{J}_k^{-1} = - \mathbf{J}_k$, and
\begin{equation}
    \mathbf{R} = \mathbf{J}_1 + \mathbf{v}^T \mathbf{M}^{-1}_0 \mathbf{v}.
\end{equation}
Therefore, $\pf\mathbf{M}$ can be computed with $\mathcal{O}(N^2)$ time complexity if $\pf\mathbf{M}_0$ and $\mathbf{M}^{-1}_0$ have been computed and stored in memory. The memory complexity is also $\mathcal{O}(N^2)$ for storing $\mathbf{M}^{-1}_0$.

To update $\mathbf{M}_0^{-1}$ to $\mathbf{M}^{-1}$, one needs to utilize the Woodbury matrix identity
\begin{equation} \label{eq:pf_inv_update}
\begin{split}
    \mathbf{M}^{-1} &= (\mathbf{M}_0 - \mathbf{v} \mathbf{J}_1 \mathbf{v}^T)^{-1} \\
    &= \mathbf{M}^{-1}_0 - \mathbf{M}^{-1}_0 \mathbf{v} 
    (-\mathbf{J}_1^{-1} + \mathbf{v}^T \mathbf{M}^{-1}_0 \mathbf{v} )^{-1} 
    \mathbf{v}^T \mathbf{M}^{-1}_0 \\
    &= \mathbf{M}^{-1}_0 + (\mathbf{M}^{-1}_0 \mathbf{v}) 
    \mathbf{R}^{-1} 
    (\mathbf{M}^{-1}_0 \mathbf{v} )^T,
\end{split}
\end{equation}
where we have used $(\mathbf{M}^{-1}_0)^T = -\mathbf{M}^{-1}_0$. This update of $\mathbf{M}^{-1}$ also has $\mathcal{O}(N^2)$ complexity. 

\subsection{Rank-\texorpdfstring{$k$}{k} update}

When $\mathbf{M}$ and $\mathbf{M}_0$ are different at $j_1, j_2, ..., j_k$ rows and columns, one can make $\mathbf{v}$ an $N \times 2k$ matrix with the first $k$ columns given by the non-zero rows of $\mathbf{M} - \mathbf{M}_0$, and the last $k$ columns given by $\mathbf{e}_{i,n} = \delta_{i,j_n}$. Then we still have
\begin{equation}
    \mathbf{M - M}_0 = -\mathbf{v} \mathbf{J}_k \mathbf{v}^T,
\end{equation}
and
\begin{equation}
    \mathbf{R} = \mathbf{J}_k + \mathbf{v}^T \mathbf{M}^{-1}_0 \mathbf{v},
\end{equation}
\begin{equation} \label{eqn: LRU_k}
    \frac{\pf \mathbf{M}}{\pf\mathbf{M}_0} = \frac{\pf \mathbf{R}}{\pf \mathbf{J}_k},
\end{equation}
\begin{equation}
    \mathbf{M}^{-1} = \mathbf{M}^{-1}_0 + (\mathbf{M}^{-1}_0 \mathbf{v}) 
    \mathbf{R}^{-1} 
    (\mathbf{M}^{-1}_0 \mathbf{v} )^T,
\end{equation}
where $\mathbf{R}$ is a $2k \times 2k$ matrix. The overall time complexity is $\mathcal{O}(k N^2)$ in the typical $k \ll N_e$ case, and the memory complexity is still $\mathcal{O}(N^2)$.

\subsection{LRU for HFPS}

Combining the HFPS wave-function in Eq.\,\eqref{eqn: hfps} and the block Pfaffian formula in Eq.\,\eqref{eq:pf_block}, one obtains
\begin{equation}
\begin{split}
    & \psi(\mathbf{n}) = J(\mathbf{n})\times \pf(\mathbf{n} \star \mathbf{F}^{vv} \star \mathbf{n}) \\
    &\times \pf[\tilde{\mathbf{F}}^{hh} + (\tilde{\mathbf{F}}^{vh}(\mathbf{n})^T \star \mathbf{n}) (\mathbf{n} \star \mathbf{F}^{vv} \star \mathbf{n})^{-1} (\mathbf{n} \star \tilde{\mathbf{F}}^{vh}(\mathbf{n}))].
\end{split} 
\end{equation}
$J(\mathbf{n})$ cannot be computed by local updates, but it does not involve the Pfaffian complexity. The first Pfaffian $\pf(\mathbf{n} \star \mathbf{F}^{vv} \star \mathbf{n})$ can be computed by LRU in Eq.\,\eqref{eqn: LRU_k}. In LRU, one also keeps track of the matrix inverse in Eq.\,\eqref{eq:pf_inv_update}, so $\mathbf{M}^{-1} = (\mathbf{n} \star \mathbf{F}^{vv} \star \mathbf{n})^{-1}$ doesn't need to be recomputed from scratch. Then the complexity of the second Pfaffian is only $\mathcal{O}(\tilde{N}^3 + N^2 \tilde{N})$. In the typical $\tilde{N} < N$ case, the overall complexity of LRU is $\mathcal{O}(N^2 \tilde{N})$ for HFPS, assuming that the ANN complexity does not exceed $\mathcal{O}(N^2)$. Therefore, LRU provides an efficient approach to compute HFPS wave-functions in VMC.

\section{Symmetry} \label{sec: symmetry}

\subsection{Symmetry in fermion systems}


For simplicity, we focus on spinless fermions in this subsection, while the discussion can be directly generalized to spinful fermions. We consider a group of symmetry operations $g \in G$ over which a fermion Hamiltonian is invariant. This group can include translations, as well as point group operations, which include rotations, reflections, inversions, etc. 
A symmetry operator acts on a fermion as follows
\begin{equation}
    \hat g \hat c_{\mathbf{x}}^\dagger \hat g^{-1} = \hat c_{g \mathbf{x}}^\dagger,
\end{equation}
where $g {\bf x} = {\bf x'}$ applies a symmetry operation to the vector ${\bf x}$ to produce a new vector ${\bf x'}$. The action of  ${\hat g}$ on a Fock state is given by
\begin{equation}
    {\hat g} \ket{\mathbf{n}} 
    = \prod_i^N \hat c_{g \mathbf{x}_i}^\dagger \ket{0}
    = \Pi(\mathbf{n}, g) \ket{g\mathbf{n}},
\end{equation}
where $g \mathbf{n}$ describes the new occupation numbers after operation $g$, and $\Pi(\mathbf{n}, g) = \pm 1$ is the additional sign associated with fermion permutations. For instance, we consider a translation by one lattice site for a 1D system with PBC,
\begin{equation}
    \hat T_1 \ket{0, 1, 0, 1} = \hat T_1 \hat c_{2}^\dagger \hat c_{4}^\dagger \ket{0}
    = \hat c_{3}^\dagger \hat c_{1}^\dagger \ket{0}
    = -\ket{1, 0, 1, 0},
\end{equation}
in which case $\Pi(\mathbf{n}, T_1) = - 1$.

Given an arbitrary state $\ket{\psi}$, one can obtain the symmetrized state
\begin{equation}
    \ket{\psi^\mathrm{symm}} = \sum_{g \in G} \chi_g^* \, {\hat g} \ket{\psi},
\end{equation}
where $\chi_g$ is the character of the symmetry representation. 
The wave-function component is given by
\begin{equation} \label{eqn: trans_symm}
\begin{split}
    \psi^\mathrm{symm}(\mathbf{n}) 
    &= \sum_g \chi_g^* \braket{\mathbf{n}|\hat g|\psi} \\
    &= \sum_g \Pi(\mathbf{n}, g) {\chi_g} \psi(g \mathbf{n}).
\end{split}
\end{equation}
Therefore, a direct computation of symmetry projected state $\psi^\mathrm{symm}(\mathbf{n})$ requires $\psi(g \mathbf{n})$ for all group elements $g$, leading to an additional $\mathcal{O}(M)$ complexity as the translation group is extensive in system size. As we will show, we can reduce the number of elements needed for symmetry projection to $\mathcal{O}(1)$, by intelligently choosing our Pfaffian couplings.  

\subsection{Translation symmetry in HFPS}

Consider the translated HFPS wave-function
\begin{equation} 
\begin{split}
    \psi(T_\mathbf{y}\mathbf{n}) &=
    J(T_{\bf y} {\bf n} ) \\ 
    \times & \mathrm{pf} \begin{pmatrix} 
       (T_{\bf y} {\bf n}) \star \mathbf{F}^{vv} \star (T_{\bf y} {\bf n}) & (T_{\bf y} {\bf n}) \star \tilde{\mathbf{F}}^{vh} (T_{\bf y} {\bf n})\\
       - \tilde{\mathbf{F}}^{vh} (T_{\bf y} {\bf n})^T \star (T_{\bf y} {\bf n}) & \tilde{\mathbf{F}}^{hh} 
    \end{pmatrix},
\end{split}
\end{equation}
where we have three $\mathbf{y}$ dependent parts, namely $J(T_{\bf y} {\bf n} )$, $(T_{\bf y} {\bf n}) \star \mathbf{F}^{vv} \star (T_{\bf y} {\bf n})$, and $(T_{\bf y} {\bf n}) \star \tilde{\mathbf{F}}^{vh} (T_{\bf y} {\bf n})$.

As shown in Fig.\,\ref{fig: HFPS_translation}, the CNN output for the Jastrow factor is an array $\mathcal{J}$, and for each translation $\mathbf{y}$ we only select the element at position $\mathbf{0}$ of the output array, then
\begin{equation}
    J(T_\mathbf{y} \mathbf{n}) = \mathcal{J}_\mathbf{0}(T_\mathbf{y} \mathbf{n})
    = \mathcal{J}_\mathbf{-y}(\mathbf{n}),
\end{equation}
where we utilize the translation equivariance of CNN in the last step. Therefore, we can avoid recomputing $J(T_\mathbf{y}\mathbf{n})$ for different $\mathbf{y}$ by obtaining the full array $\mathcal{J}(\mathbf{n})$ from CNN in a single forward pass. For the second part, we have
\begin{equation}
    (T_{\bf y} {\bf n}) \star F_{\mathbf{x}, \mathbf{x}'}^{vv} \star (T_{\bf y} {\bf n})
    \simeq  \mathbf{n} \star F_{\mathbf{x}+\mathbf{y}, \mathbf{x}'+\mathbf{y}}^{vv} \star \mathbf{n},
\end{equation}
where we have used $\simeq$ to denote that the matrices are equal up to a permutation of rows and columns, which produces a tractable fermion sign, $\Pi(\mathbf{n}, T_{\bf y})$. Similarly, we also have
\begin{equation}
    (T_{\bf y} {\bf n}) \star \tilde{F}_{\mathbf{x},\tilde{p}}^{vh} (T_{\bf y} {\bf n})
    = (T_{\bf y} {\bf n}) \star \tilde{F}_{\mathbf{x}-\mathbf{y},\tilde{p}}^{vh} ({\bf n})
    \simeq \mathbf{n} \star \tilde{F}_{\mathbf{x},\tilde{p}}^{vh} ({\bf n}),
\end{equation}
where again the output is equal up to a permutation of rows. A similar relation was also utilized in NNBF to avoid recomputing CNN outputs for different translation $\mathbf{y}$~\cite{Romero_CP25_NNBF}.

Therefore, the translated wave-function becomes
\begin{equation}
\begin{split}
\psi(T_\mathbf{y}\mathbf{n}) &= \Pi(\mathbf{n}, T_{\bf y}) \times 
   \mathcal{J}_{-\mathbf{y}}({\bf n} ) \\ 
   & \times \pf \begin{pmatrix} 
        {\bf n} \star F^{vv}_{{\bf x - \bf y},{\bf x' - \bf y}  } \star {\bf n} & \mathbf{n} \star \tilde{F}_{\mathbf{x},\tilde{p}}^{vh} ({\bf n}) \\
       - \tilde{F}_{\mathbf{x},\tilde{p}}^{vh} ({\bf n})^T \star \mathbf{n} & \tilde{F}^{hh}_{\tilde p, \tilde q} 
    \end{pmatrix},
\end{split}
\end{equation}
where one still has to recompute the Pfaffian values for different translations $\mathbf{y}$ in general. Alternatively, one can enforce $F^{vv}_{{\bf x - \bf y},{\bf x' - \bf y} } = F^{vv}_{{\bf x},{\bf x'} }$ to remove the $\mathbf{y}$ dependence of the Pfaffian matrix, but this constraint severely restricts the expressivity of the mean-field part $F^{vv}$, and therefore the HFPS. Sublattice structure can be introduced as a balance between expressive power and efficiency.

\subsection{Sublattice Symmetry}

In practice, we rarely choose $F^{vv}$ to be fully translationally invariant, as this inhibits our ability to model symmetry-broken phases. We instead allow symmetry breaking within an enlarged unit cell, which is informed by our prior knowledge of the problem. Typically, we choose this unit cell by first training a Slater determinant Ansatz, and then choosing our Pfaffian unit cell based on the translational symmetry of the density matrix.   

We consider a situation where our couplings between visible fermions $F_{vv}$ are invariant under some subset of the translations $\{ {\bf z} \} \subset \{ {\bf y} \}$~\cite{Misawa_CPC19_mVMC}
\begin{equation}    
F^{vv}_{{\bf x} + {\bf z}, {\bf x'} + {\bf z}} = F^{vv}_{{\bf x}, {\bf x'}  }.
\end{equation}

In order to restore full translational symmetry, we need to symmetrize over the translations within the unit cell $\{ {\bf w} \}$, where $\{ {\bf w} \} = \{ {\bf y} \} / \{ {\bf z} \}$. This group division operation indicates that the symmetry operations in $\{ {\bf y} \}$ can be formed by combining the operations in $\{ {\bf w} \}$ and $\{ {\bf z} \}$. The wave-function of our sublattice symmetric CNN is written as
\begin{equation}
\begin{split}
\psi({\bf n}) &= \sum_{\bf w} \Big[ \Pi({\bf n}, T_{\bf w}) \times \mathcal{J}'_{-\bf w}({\bf n} ) \\ 
  & \times \pf \begin{pmatrix} 
        {\bf n} \star F^{vv}_{{\bf x - \bf w},{\bf x' - \bf w}  } \star {\bf n} & {\bf n} \star F^{vh}_{{\bf x - \bf w}, {\tilde q}} ({\bf n}) \\
       - F^{vh}_{{\bf x - \bf w}, {\tilde q}} ({\bf n})^T \star {\bf n} & F^{hh}_{\tilde p \tilde q} 
    \end{pmatrix} \Big],
    \end{split}
\end{equation}
where we have defined the reduced symmetry Jastrow factor
\begin{equation}
    \mathcal{J}'_{\bf w}({\bf n}) \equiv \sum_{\mathbf{z} \in \{\mathbf{z}\}} \mathcal{J}_{\mathbf{w}+\mathbf{z}}(\mathbf{n}).
\end{equation}
As before, we only need a single CNN to get all of the features, but we have to compute $|\{ \bf w \}|$ Pfaffians, where $|\{ \bf w \}|$ is the number of sites within a unit cell.

\subsection{HFPS + GCNN} \label{sec: HFPS+GCNN}

Hidden fermion Pfaffian states can also be combined with group convolutional neural networks (GCNNs) \cite{Roth_PRB23_GCNN, Roth_arxiv21_GCNN, Cohen_arxiv16_GCNN} instead of CNNs. The GCNN generalizes the CNN from the translation group to general non-abelian groups $G$. The feature maps of a GCNN encode a representation over the full space group instead of the translation group $\mathcal{J}_\mathbf{x} \rightarrow \mathcal{J}_g$, where $g \in G$ is an element of the space group. For lattices, $G$ includes translations, reflections, rotations, and spin-parity symmetry. In a similar fashion to how the translation group is sub-divided for the CNN, we can divide the symmetry operations of the space group into two subsets; translations by vectors of the sublattice unit cell $\{ {\bf z} \}$, over which the full HFPS state is invariant, and the remaining operations, which include translations within the unit cell and point group operations. In order to make the HFPS fully invariant, we project over the remaining operations, $h \in H$, where $H = G / \{ {\bf z} \}$. Then
\begin{equation}
\begin{split}
    \psi({\bf n}) &=  \sum_{h \in H} \Big[\Pi({\bf n}, h^{-1}) \mathcal{J}'_{h}({\bf n} ) \times \\ 
    & \times \pf \begin{pmatrix} 
        {\bf n} \star F^{vv}_{h{\bf x},h {\bf x'}} \star {\bf n} & {\bf n} \star F^{vh}_{h {\bf x}, {\tilde q}} ({\bf n}) \\
       -F^{vh}_{h {\bf x}, {\tilde q}} ({\bf n})^T \star {\bf n} & F^{hh}_{\tilde p, \tilde q} 
    \end{pmatrix} \Big],
    \end{split}
\end{equation}
where $h {\bf z}$ describes the vector ${\bf z}$ transformed by operation $h$, and
\begin{equation}
    \mathcal{J}'_{h}({\bf n} ) \equiv \sum_{\mathbf{z} \in \{\mathbf{z}\}} \mathcal{J}_{h\mathbf{z}}(\mathbf{n})
\end{equation}
is the partially symmetrized Jastrow factor. 

{
\renewcommand{\arraystretch}{1.5}

\begin{table}[ht!]
    \centering
    \begin{tabular}{c|c|c|c}
      $U$ & $E_\mathrm{ED}/N_\mathrm{site}$ & $E/N_\mathrm{site}$ & $\sigma^2/N_\mathrm{site}$ \\ \hline\hline
      4 & -1.223808595 & -1.223807(7) & $8.8 \times 10^{-6}$ \\ \hline
      6 & -1.147397817 & -1.147396(7) & $1.3 \times 10^{-5}$ \\ \hline
      8 & -1.094397918 & -1.094396(4) & $2.6 \times 10^{-5}$ \\ \hline
      10 & -1.056472499 & -1.056468(9) & $6.1 \times 10^{-5}$
    \end{tabular}
    \caption{Raw data of Fig.\,\ref{fig: 4x4_FL} in the $4 \times 4$ lattice with particle density $n = 5/8$.}
    \label{tab: small scale}
\end{table}

\begin{table}[ht!]
    \centering
    \begin{tabular}{c|c|c|c}
      $U$ & $E_\mathrm{AFQMC}/N_\mathrm{site}$ & $E/N_\mathrm{site}$ & $\sigma^2/N_\mathrm{site}$ \\ \hline\hline
      -4 & -2.8603(8) & -2.859578 & $5.9 \times 10^{-3}$ \\ \hline
      -8 & -4.5258(9) & -4.525834 & $2.3 \times 10^{-5}$ \\
    \end{tabular}
    \caption{Raw data of Fig.\,\ref{fig: half-filling} in the $8 \times 8$ lattice at half filling. The HFPS data shown here is the one with sublattices.}
    \label{tab: half filling}
\end{table}

\begin{table}[ht!]
    \centering
    \begin{tabular}{c|c|c|c}
      System size & $E/N_\mathrm{site}$ & $\sigma^2/N_\mathrm{site}$ & Training Time / h \\ \hline\hline
      $4 \times 4$ & -0.74177(5) & & \\ \hline
      $8 \times 4$ & -0.76690(1) &  0.016320 & 100 \\ \hline
      $12 \times 4$ & -0.76608(1) & 0.028784 & 200 \\ \hline
      $16 \times 4$ & -0.76413(3) & 0.031283 & 1500 \\
    \end{tabular}
    \caption{Raw data of Fig.\,\ref{fig: CDW} in the $L \times 4$ lattice with 1/8 doping and $U=8$. Training time is listed in hours on NVIDIA H100 80GB GPUs.}
    \label{tab: stripe energies}
\end{table}
}

\section{Raw data} \label{sec: raw data}

In Table \ref{tab: small scale}-\ref{tab: stripe energies}, we include the raw data of our variational simulations.

\clearpage

\bibliography{reference}

\end{document}